\documentclass[pra,twocolumn,floatfix,superscriptaddress,longbibliography]{revtex4-1}

\usepackage{amssymb,amsmath,amstext}
\usepackage{mathtools}
\usepackage{graphicx}
\usepackage{epstopdf}
\usepackage{color}
\usepackage{rotating}
\usepackage{bm}
\usepackage{slashed}
\usepackage{appendix}
\usepackage[T1]{fontenc}
\usepackage{bbold}
\usepackage{bbm}
\usepackage{latexsym}
\usepackage[colorlinks=true,citecolor=blue,linkcolor=magenta]{hyperref}

\usepackage[final]{showlabels}

\newcommand{\mN}{\mathcal N}
\newcommand{\mO}{\mathcal O}

\newcommand{\bra}[1]{\mbox{$\langle #1 |$}}
\newcommand{\ket}[1]{\mbox{$| #1 \rangle$}}

\newcommand{\gs}{\ket{\mathrm{GS}}}

\begin{document}

\title{Noise amplification at spin-glass bottlenecks of quantum annealing:\\ a solvable model}

\author{David Roberts}

\affiliation{Department of Physics,  University  of  Chicago,  Chicago,  Illinois  60637,  USA}
\affiliation{NASA Ames Research Center Quantum Artificial Intelligence Laboratory (QuAIL), Moffett Field, CA 94035, USA}
\affiliation{Theoretical Division, Los Alamos National Laboratory, Los Alamos, NM 87545, USA}

\author{Lukasz Cincio}
\affiliation{Theoretical Division, Los Alamos National Laboratory, Los Alamos, NM 87545, USA}

\author{Avadh Saxena}
\affiliation{Theoretical Division, Los Alamos National Laboratory, Los Alamos, NM 87545, USA}

\author{Andre Petukhov}
\affiliation{NASA Ames Research Center Quantum Artificial Intelligence Laboratory (QuAIL), Moffett Field, CA 94035, USA}
\affiliation{Google, Venice, CA 90291, USA}

\author{Sergey Knysh}
\affiliation{NASA Ames Research Center Quantum Artificial Intelligence Laboratory (QuAIL), Moffett Field, CA 94035, USA}

\affiliation{Google, Venice, CA 90291, USA}
\affiliation{SGT Inc., 7701 Greenbelt Rd., Greenbelt, MD 20770}

\begin{abstract}
To gain better insight into the complexity theory of quantum annealing,
we propose and solve a class of spin systems which contain bottlenecks of the kind expected to dominate the runtime of
quantum annealing as it tries to solve difficult optimization problems. We uncover a noise amplification effect at these bottlenecks, whereby tunneling rates caused by flux-qubit noise scale in proportion to
the number of qubits $\mN$ in the limit that $\mN\to \infty$. By solving the incoherent annealing dynamics exactly, we find a wide range of regimes where the probability that a quantum annealer remains in the ground-state upon exiting the bottleneck is close to one-half. We corroborate our analysis with detailed simulations of the performance of the D-Wave 2X quantum annealer on our class of computational problems.
\end{abstract}
\maketitle

\section{Introduction} 

Quantum Annealing (QA), a quantum heuristic for approximately solving NP-hard
binary optimization problems, is already in commercial use
\cite{boixo_evidence_2014,king_observation_2018,venturelli_reverse_2019,venturelli_compiling_2018,neukart_traffic_2017,rosenberg_solving_2016,rieffel_case_2015} in machine learning and
artificial intelligence applications. The algorithm
works by mapping Quadratic Unconstrained Binary Optimization (QUBO) problems
to the problem of solving for the ground state of a spin glass Hamiltonian. The
time-complexity of QA, however, that is, how the required resources for running
the algorithm scale with problem size, is still under investigation. The
scaling behavior has been computed only for several optimization problems
\cite{knysh_computational_2015, young_first-order_2010, morita_convergence_2006, santoro_theory_2002}.

\begin{figure}[t!]
        \includegraphics[width=\columnwidth]{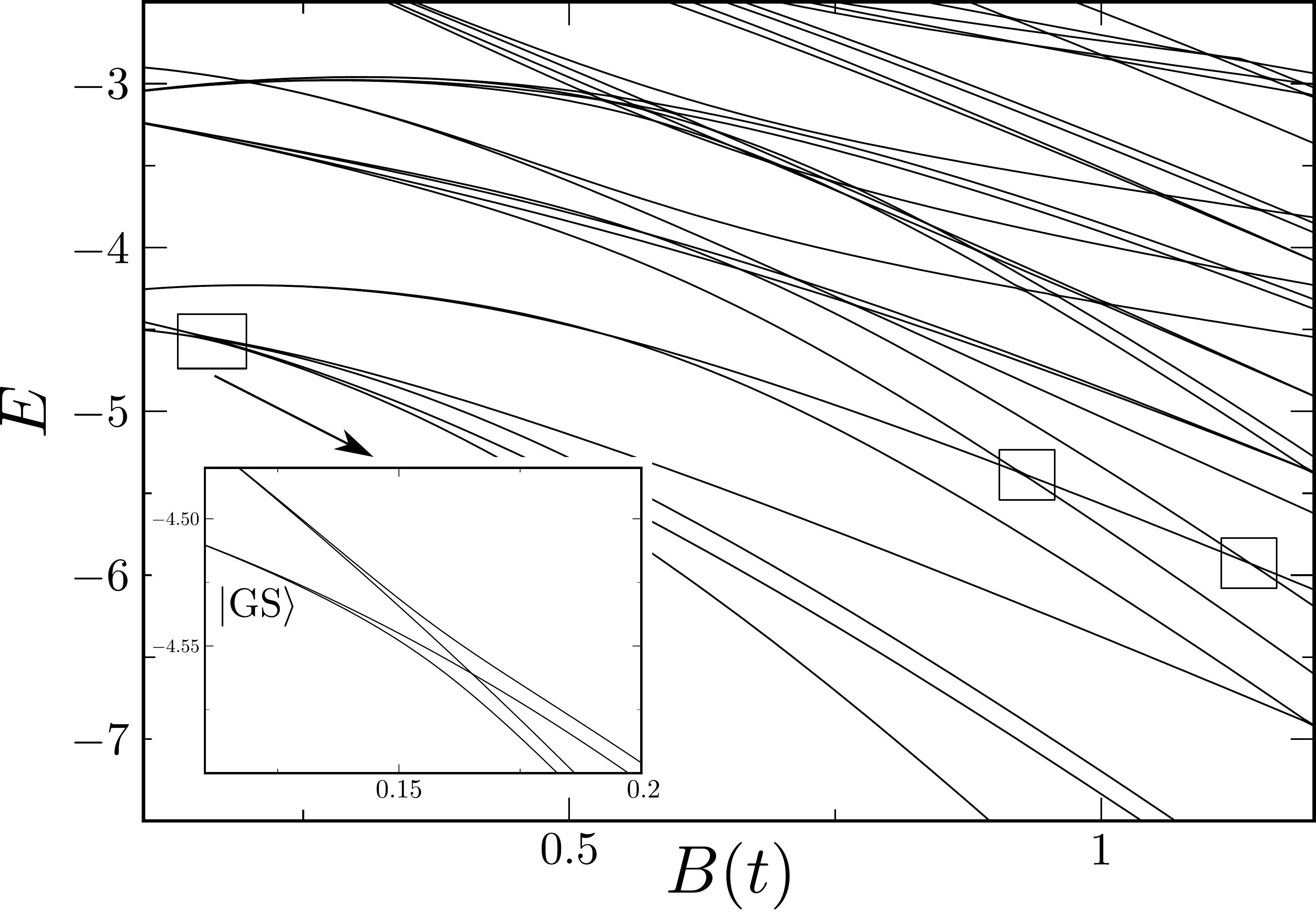}
        \caption{ One-dimensional bottlenecks of quantum annealing. The plot shows some exponentially-small gaps in the spectrum (obtained by exact diagonalization) of a 1+1 dimensional $\mN=7$ transverse-field Ising spin glass. Only one is a bottleneck (i.e. involves the instantaneous ground state $\gs$, shown in the inset). Multiqubit-tunneling at this bottleneck is asymptotically proportional to the number of qubits~$\mN$, magnifying annealing errors in the adiabatic limit. The figure shows instantaneous eigenenergies $E$ of the Hamiltonian in Eq. \eqref{eq:QA_ham} as a function of $B(t)$ for representative values of $J_j$ couplings in~Eq. \eqref{eq:Jj}.}
        \label{fig:low-energy}
\end{figure}

A key benchmarking problem in QA is the question of how an adiabatic quantum
computer performs on spin glass bottlenecks. Those are time intervals
during the annealing schedule where the gap shrinks exponentially with
problem size, see Fig. \ref{fig:low-energy}. A cascade of hard bottlenecks
was found \cite{santoro_theory_2002} in the ordered phase of the annealing
process. Over a decade later, another work \cite{knysh_computational_2015}
provided an exactly solvable spin glass system, where the scaling of the
gap at these bottlenecks was obtained via analytical arguments, as opposed
to the usual, numerical treatment. As a result, it is now understood that
the time-complexity of quantum annealing for large problems is dominated by
spin glass bottlenecks \cite{lanting_experimental_2017,king_degeneracy_2016,steiger_heavy_2015,boixo_evidence_2014,dickson_thermally_2013,altshuler_anderson_2010}.
There are two general features of these bottlenecks, (i) an exponentially small
gap as the system size $\mN$ grows, and (ii) a quantum tunneling event that
flips $\mO(\mN)$ spins. 

These features can be embedded in a simple model,
which can then be solved exactly (some properties are accurately accessible
only in the asymptotic limit of $\mN \rightarrow \infty$). To the best of our
knowledge, there have been no analytical studies of the effects of realistic
(i.e. longitudinal) qubit noise at bottlenecks of QA in the presence
of frustration, in the limit of large $\mN$ \cite{keck_dissipation_2017,
deng_decoherence_2013}. In this study, we ultimately find that, at an annealing bottleneck, the effects
of frustration on multiqubit tunneling are washed-out in the large-$\mN$ limit,
leaving behind a large-$\mN$ noise amplification effect. In particular, we
find that tunneling rates in a wide class of frustrated spin chains diverge
as $\mN M^2$, where $M$ is a suitably-defined bulk
spontaneous magnetization. This gives analytical confirmation of prior work, which has found that the effective multiqubit noise spectral density at an annealing bottleneck
grows as the multiqubit Hamming distance between the crossing states \cite{boixo_computational_2016, amin_role_2009}.\\

The rest of the paper is organized as follows. We introduce and discuss
basic properties of our exactly solvable model of bottlenecks in QA in
Section~\ref{sec:model}. Section~\ref{sec:noise} concentrates on the
effects of ambient noise at the QA bottlenecks. We solve the Redfield equation to predict the behavior of the annealing processor at the
bottleneck at finite temperature in Section~\ref{sec:largeN}. We conclude with the main results in Section~\ref{sec:discussion}.

\section{A minimal model of a spin glass bottleneck of QA}
\label{sec:model}
\begin{figure} 
        \includegraphics[width=0.85\columnwidth]{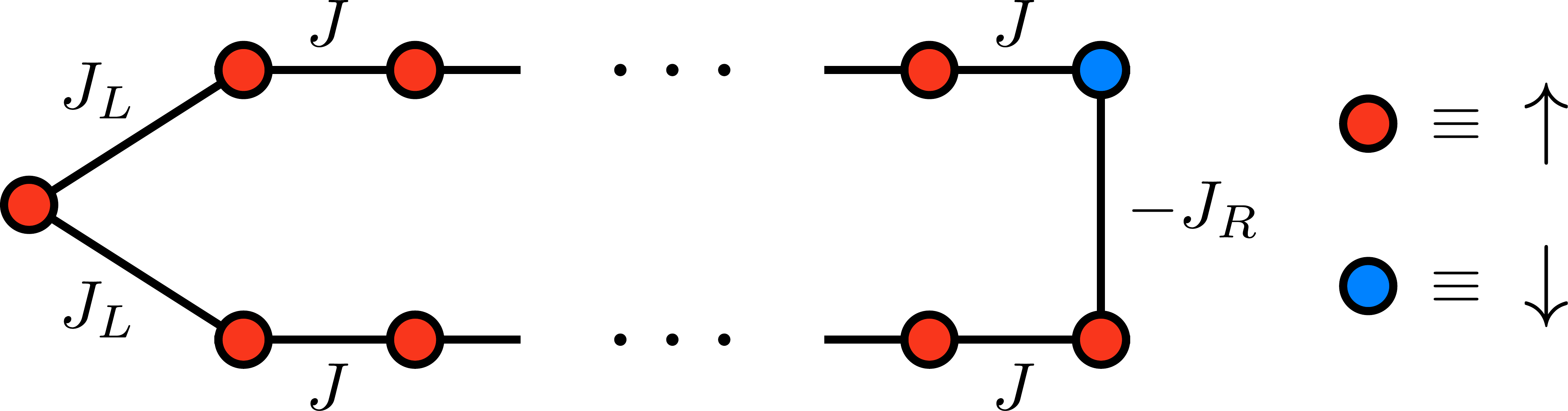}
        \caption{The Frustrated Ring (c.f. Eq. \eqref{eq:fmodel}) is specified by the above weighted graph. Links correspond to couplings between sites represented by circles. It is a solvable 1D model of a spin glass bottleneck of QA. Red (blue) circle denotes spin up (down).}
        \label{fig:fmodel}
\end{figure}

We now search for the simplest class of Ising spin glasses containing a {\it spin glass bottleneck}, where the minimum gap decreases exponentially as the number of qubits $\mN\to \infty$.  Note that spin glass bottlenecks with frustration are impossible to realize with mean-field-like problem Hamiltonians (i.e. problem Hamiltonians with all-to-all interactions), thus motivating an investigation of the one-dimensional case. Defining Pauli matrices $\sigma^\alpha_j$, $\alpha=x,z$ acting on sites $j$ of a lattice, the one-dimensional transverse-field Ising spin glass has Hamiltonian
\begin{align}
    \hat{H}_0(t) &\equiv \hat{H}_P+B(t)\sum_{j} \hat{\sigma}^x_j.\label{eq:QA_ham}
\end{align}
The parameter $B(t)$ here represents a uniform transverse magnetic field, and the problem Hamiltonian $\hat{H}_P$ for the one-dimensional Ising spin glass is\\
\begin{align}
\hat{H}_P&\equiv -\sum_{j=1}^\mN J_j \sigma^z_j\sigma^z_{j+1} ,\label{eq:fmodel}
\end{align}
where here, $J_j$ denotes a coupler connecting qubits $j$ and $j+1$, and it is understood that the qubit indices $j$ are to be interpreted modulo $\mN$. The one-dimensional transverse-field Ising model can be mapped to free fermions \cite{lieb_two_1961}, and is thus a good place to gain some useful physical intuition about these bottlenecks. 
 The simplest one-dimensional annealing problem containing a spin glass bottleneck is the Frustrated Ring, the one-dimensional spin system depicted in Figure \ref{fig:fmodel}. The couplings $J_j$
for the Frustrated Ring are given as follows:
\begin{equation} \label{eq:Jj}
J_j =
\begin{cases} 
      J_L  & j = n, n + 1 \\
      -J_R & j = 2n+1 \\
      J  & \text{otherwise}
\end{cases} \ ,
\end{equation}
here, $0<J_R<J_L<J$, and we make the total number of qubits $\mN\equiv 2n+1$ odd, to make the problem more symmetric (and thus more amenable to an exact calculation).  The Frustrated Ring is a minimal model of a spin glass bottleneck, because, in one dimension, one must modify at least three couplers in an otherwise uniform graph to achieve a spin glass bottleneck; the Frustrated Ring saturates this lower bound. 

\begin{figure}
    \centering
    \includegraphics[width = 0.9 \columnwidth]{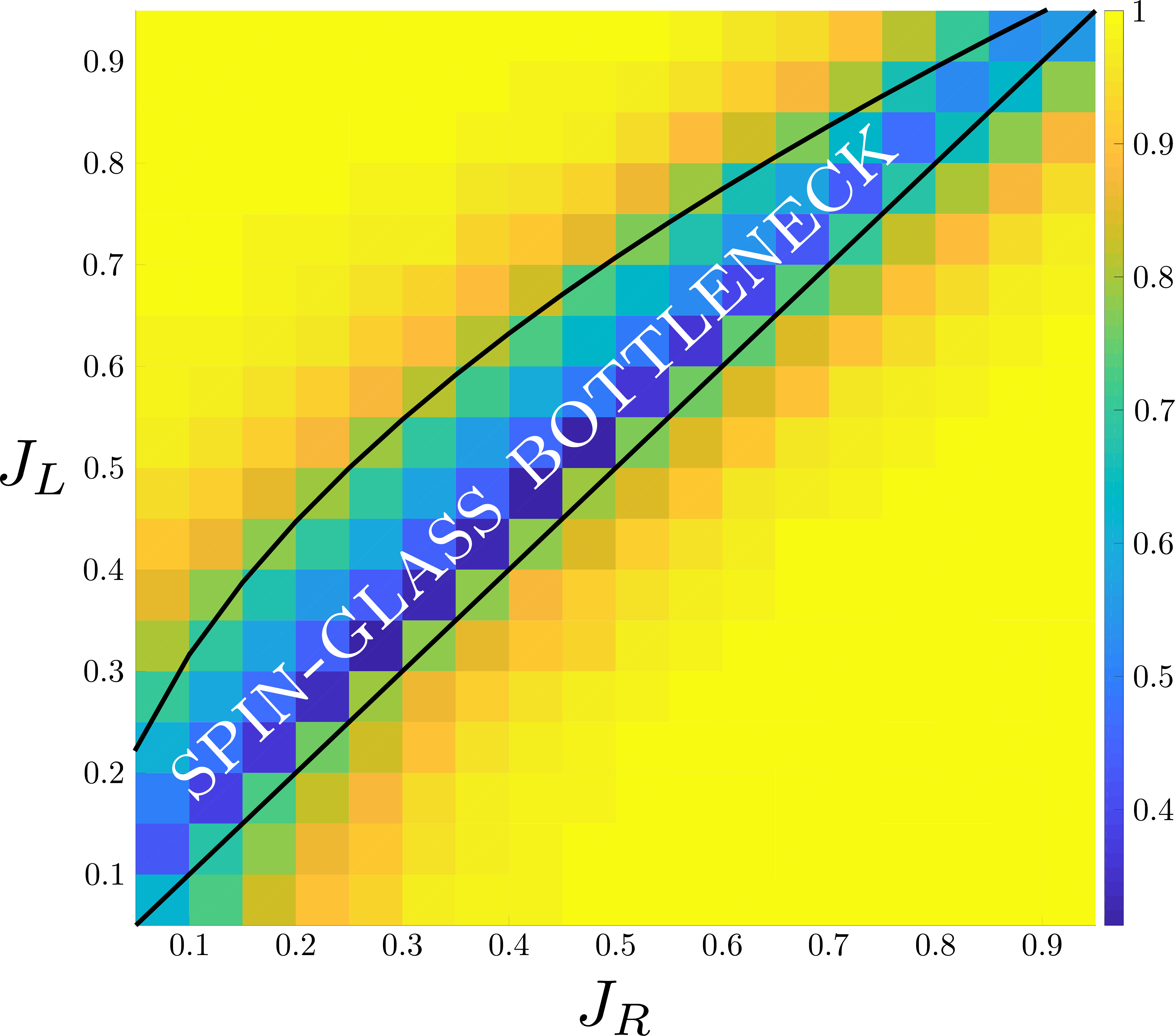}
    \caption{Performance of the D-Wave 2X quantum annealer on the Frustrated Ring spin-glass benchmark (c.f. Figure \ref{fig:fmodel}). In the regime $0<J_R<J_L<J$,~$JJ_R>J_L^2$ (the region between the two black curves), the annealing schedule of the Frustrated Ring contains a spin-glass bottleneck. Within the bottleneck regime, the probability that the D-Wave 2X returns the groundstate of the Ising spin glass in Eq. \eqref{eq:fmodel} decreases noticeably from $\sim 1$ (yellow) to $\sim 0.4$ (dark blue). Here, $\mN =8$, and the coupling $J\equiv 1$ in the bulk of the chain.}
    \label{fig:scan}
\end{figure}

\begin{figure} 
        \includegraphics[width=\columnwidth]{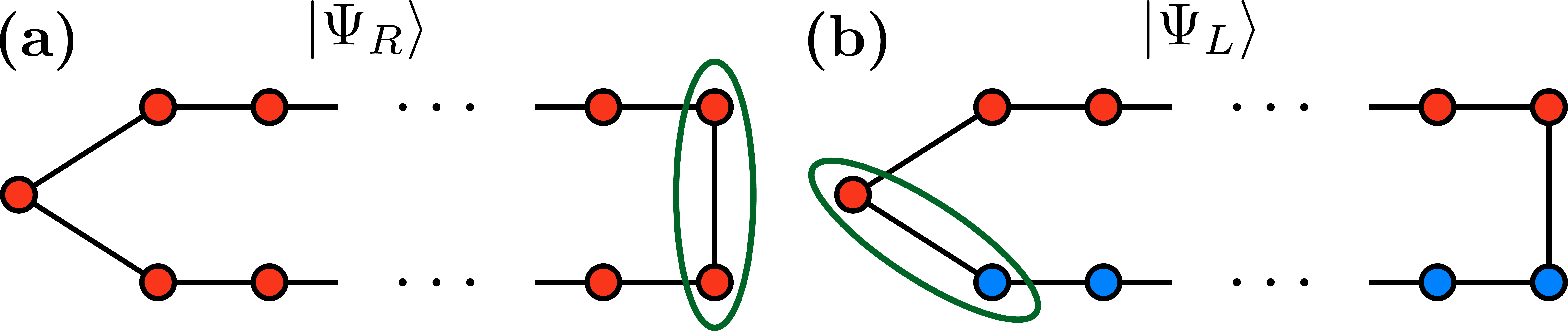}
        \caption{Lowest energy states of the Frustrated Ring, at the end of the annealing process. The green loop denotes a bond that gives a positive contribution to the energy of the Hamiltonian in Eq. \eqref{eq:fmodel}. In {\bf (a)}, $\ket{\Psi_R}$ is formed by violating the anti-ferromagnetic bond $J_R$. $\ket{\Psi_L}$ is obtained by violating either of the two weak ferromagnetic bonds $J_L$, as shown in {\bf (b)}.}
        \label{fig:states}
\end{figure}

\begin{figure}
        \includegraphics[width=\columnwidth]{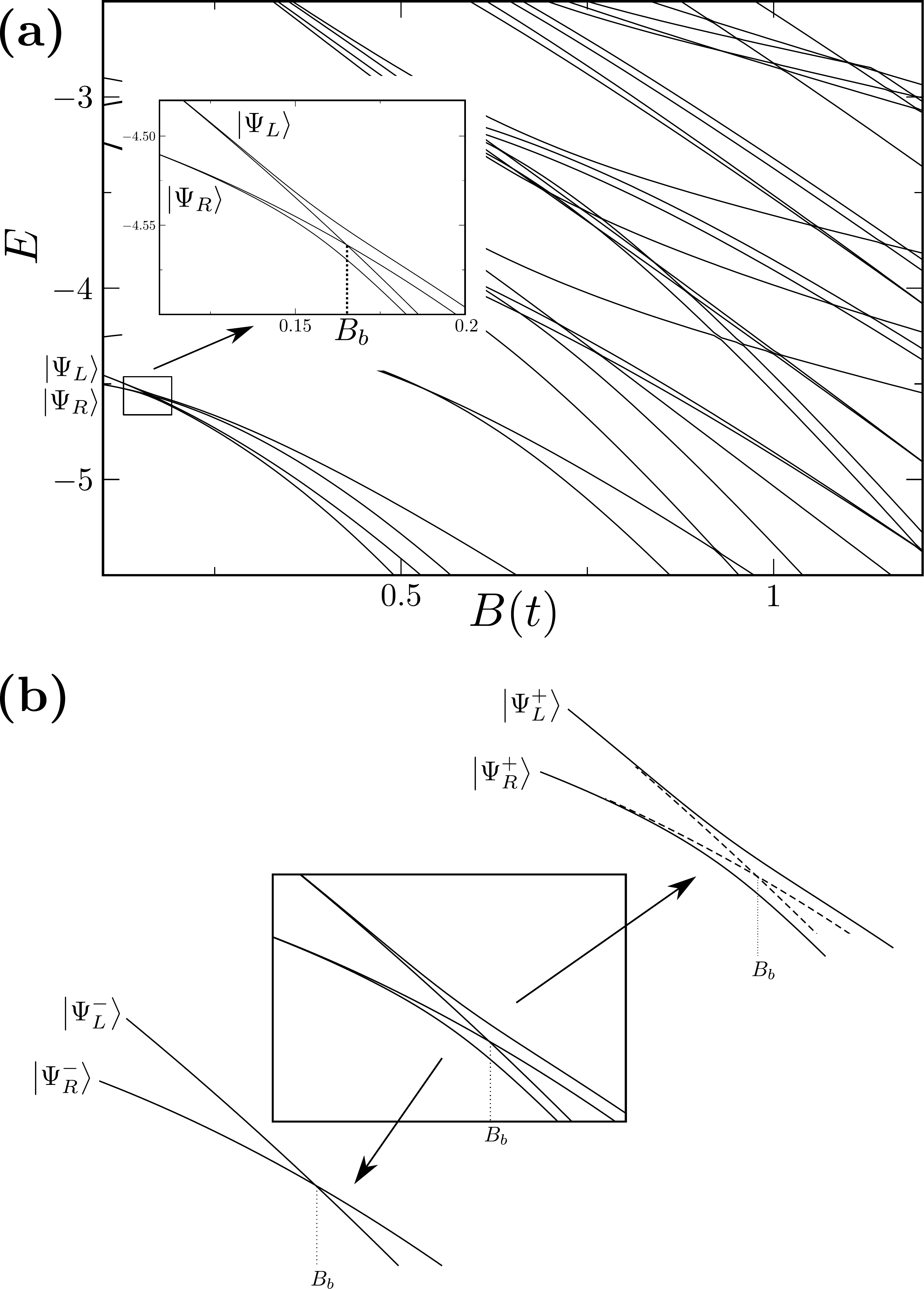}
        \caption{{\bf (a)} The low-energy spectrum of an $\mN=7$ Frustrated Ring computed for representative values of couplings $J_j$ in Eq. \eqref{eq:QA_ham} and Eq. \eqref{eq:Jj}. Note the avoided crossing. {\bf (b)} Due to $\mathbb{Z}_2$ Ising symmetry, the spin glass bottleneck consists of two crossings, one occurring in each symmetry sector. In the Frustrated Ring, the odd-parity crossing is gapless. However, only the even-parity crossing is seen by the coherent annealing dynamics (see \cite{dziarmaga_dynamics_2005} for details), and thus the even-parity gap (see Eq. \eqref{eq:rigorous_gap_scaling}) determines the QA time-complexity of this problem.}
        \label{fig:lowenergy}
\end{figure}

The Frustrated Ring is frustrated, and therefore has a forced excitation in its groundstate. At zero transverse field $B(t)\equiv 0$, there are two generic positions where this excitation likes to reside: (i) at the antiferromagnetic coupler $J_R$ (forming the
frustrated groundstate $\ket{\Psi_R}$), and (ii)  at either of the
two weak ferromagnetic couplers $J_L$ (forming degenerate first-excited states $\ket{\Psi_L}$),
see Fig.~\ref{fig:states}. At a special value of the transverse-field $B_b$ within
the ordered phase of the anneal, the $\ket{\Psi_R}$ states and a pair of $\ket{\Psi_L}$ states form an avoided
crossing with a gap that scales as
\begin{align}
    \Delta_\text{min}&\underset{\mN\to\infty}{\propto} \bigg(\frac{J(J_L^2-J_R^2)}{J_R(J^2-J_L^2)}\bigg)^\mN,\label{eq:gapscaling}
\end{align}
which is exponentially small in $\mN$, as desired (see Fig.~\ref{fig:lowenergy}). Representative performance of the D-Wave 2X quantum annealer in the bottleneck regime, $JJ_R>J_L^2$,  is depicted in Fig. \ref{fig:scan}. Since $\ket{\Psi_L}$ and $\ket{\Psi_R}$ differ by flipping half of the spins,
the avoided crossing of these states has two key features in common with
spin glass bottlenecks: (i) an exponentially small gap, and (ii) a quantum
tunneling event that flips $\mO(\mN)$ spins.

\subsection{Annealing schedule}
We now analyze the annealing schedule of the Frustrated Ring, and demonstrate
that it runs in exponential time (assuming completely coherent processor dynamics). The first step in the analysis consists
of reinterpreting the system of interacting spins as a system
of non-interacting fermionic excitations (which represent dressed domain-walls in the spin representation). In this new
description, the global spin-flip symmetry  becomes a symmetry $(-1)^{N_F}$, which counts
the number of fermionic excitations modulo $2$:
\begin{align}
   (-1)^{N_F}\equiv \prod_{j=1}^\mN \hat{\sigma}^x_j \ ,~~~~~[\hat{H}_0(t),(-1)^{N_F}]=0 \ .\label{eq:parity_is_conserved}
\end{align}
The appropriate fermionic operators $\hat{\gamma}_1,\hat{\gamma}_2,\cdots,\hat{\gamma}_{2\mN}$ are Majorana fermions, i.e. they generate the Clifford algebra 
\begin{align}
    \hat{\gamma}_i\hat{\gamma}_j+\hat{\gamma}_j\hat{\gamma}_i\equiv 2\delta_{ij}.
\end{align}
These Majorana fermions are written in terms of the original spin operators via the Jordan-Wigner
transformation
\begin{eqnarray}
\hat{\sigma}^x_j &=& -i \hat{\gamma}_{2j-1}\hat{\gamma}_{2j} \ , \nonumber\\
\hat{\sigma}^z_j &=& (-i)^{j-1} \hat{\gamma}_1 \cdots \hat{\gamma}_{2j-1} \ .
\end{eqnarray}
In terms of the fermionic operators, the theory decouples into two free theories:
\begin{equation} \label{eq:HFmodel}
\begin{split}
\hat{H}^\pm_0(t) = & -i \sum_j J_j^\pm\, \hat{\gamma}_{2j}\hat{\gamma}_{2j+1}
\\ & +i B(t)\sum_j\hat{\gamma}_{2j-1}\hat{\gamma}_{2j} \ ,
\end{split}
\end{equation}
where $J^\pm_j\equiv J_j$ for $j\neq \mN$; $J_{\mN}^\pm \equiv \pm J_{\mN}$,
and $\hat{H}^\pm_0(t)$ denotes the annealing Hamiltonian $\hat{H}_0(t)$ restricted to the sector with an even and odd number of Majorana fermions, respectively. The odd-parity Hamiltonian $\hat{H}^-_0(t)$ is simplest to analyze, as all of its coupling constants are positive in the limit $B\to 0$, and thus all fermionic excitations (for sufficiently small transverse field) have positive energy. For $B>0$, the energies $\epsilon>0$ of these excitations will in general have some non-trivial dependence on the transverse field, leading to an {\it energy dispersion curve} $\epsilon(B)$. When any one of these dispersion curves crosses the horizontal line $\epsilon(B)\equiv 0$, the model becomes gapless. We call the smallest transverse field at which this happens $B_c$, i.e. the {\it critical value} of the transverse-field.

We then define $B(t)<B_c$ to be the {\it spin glass phase} of the anneal, and the remaining regime $B(t)>B_c$ to be the {\it paramagnetic phase} of the anneal. In the spin-glass phase, the Bogoliubov de-Gennes (BdG) equations
centered on the bond $J_\mN^- \equiv J_R$ may be solved exactly, producing
a fermion bound state $\hat{c}_R$ exponentially localized at the right-end of the
graph, creating the odd-parity ground state.
\begin{align}
|\Psi_R^-\rangle :=\hat{c}_{R,-}^\dag |\Omega_-\rangle.\label{eq:odd0}
\end{align}
Similarly, in the limit $\mN \rightarrow \infty$,
the BdG equations centered on the bond $J_L$ may also be solved exactly, see
Appendix~\ref{app:low}. They yield also a low-energy bound state $\hat{c}_L$ exponentially
localized at the left-end of the graph, so that the next-lowest energy state with odd-parity is
\begin{align}
    |\Psi_L^-\rangle :=\hat{c}_{L,-}^\dag |\Omega_-\rangle.\label{eq:odd1}
\end{align}
The corresponding energies of the states Eqs. (\ref{eq:odd0}-\ref{eq:odd1}) are $\epsilon_{vac}^- + \epsilon_{R}^-$, and $\epsilon_{vac}^- + \epsilon_{L}^-$, respectively, where $\epsilon_{vac}^-$ is the energy of the odd-parity vacuum state $|\Omega^-\rangle$. From the definition, it is then clear that, whenever the energy dispersion curves $\epsilon_{R}^-(B)$ and $\epsilon_{L}^-(B)$ cross, the corresponding spin-chain eigenstates Eqs. (\ref{eq:odd0}-\ref{eq:odd1}) cross in energy.

In the even-parity sector, a similar picture emerges, although the energy considerations are complicated by the fact that some defects have negative energy in the spin glass phase, as the even-parity Hamiltonian $\hat{H}^+_0$ contains a coupling constant $J_\mN^+ \equiv -J_R$ which is negative. However, after some book-keeping, one can deduce the two-lowest energy states in the even-fermion sector, which we list below:\\
\begin{align}
    |\Psi_R^+\rangle &:= |\Omega_+\rangle,\label{eq:even0}\\
    |\Psi_L^+\rangle &:=\hat{c}_{R,+}^\dag \hat{c}_{L,+}^\dag |\Omega_+\rangle.\label{eq:even1}
\end{align}
As in the odd-parity case, the corresponding energies of the states Eqs. (\ref{eq:even0}-\ref{eq:even1}) are $\epsilon_{vac}^+$, and $\epsilon_{vac}^+ + \epsilon_{L}^++\epsilon_R^+$, respectively, where $\epsilon_{vac}^+$ is the energy of the even-parity vacuum state $|\Omega^+\rangle$. Again, this means that when the energy dispersion curves $|\epsilon_{R}^+(B)|$ and $\epsilon_{L}^+(B)$ cross, the corresponding spin-chain eigenstates Eqs. (\ref{eq:odd0}-\ref{eq:odd1}) cross in energy. Here, we have taken the absolute value of the energy of the right-localized defect $\hat{c}_{R,+}^\dag$, as this defect has {\it negative} energy throughout the spin glass phase $B(t)<B_c$ of the anneal.

\begin{figure}
\includegraphics[width=0.85\columnwidth]{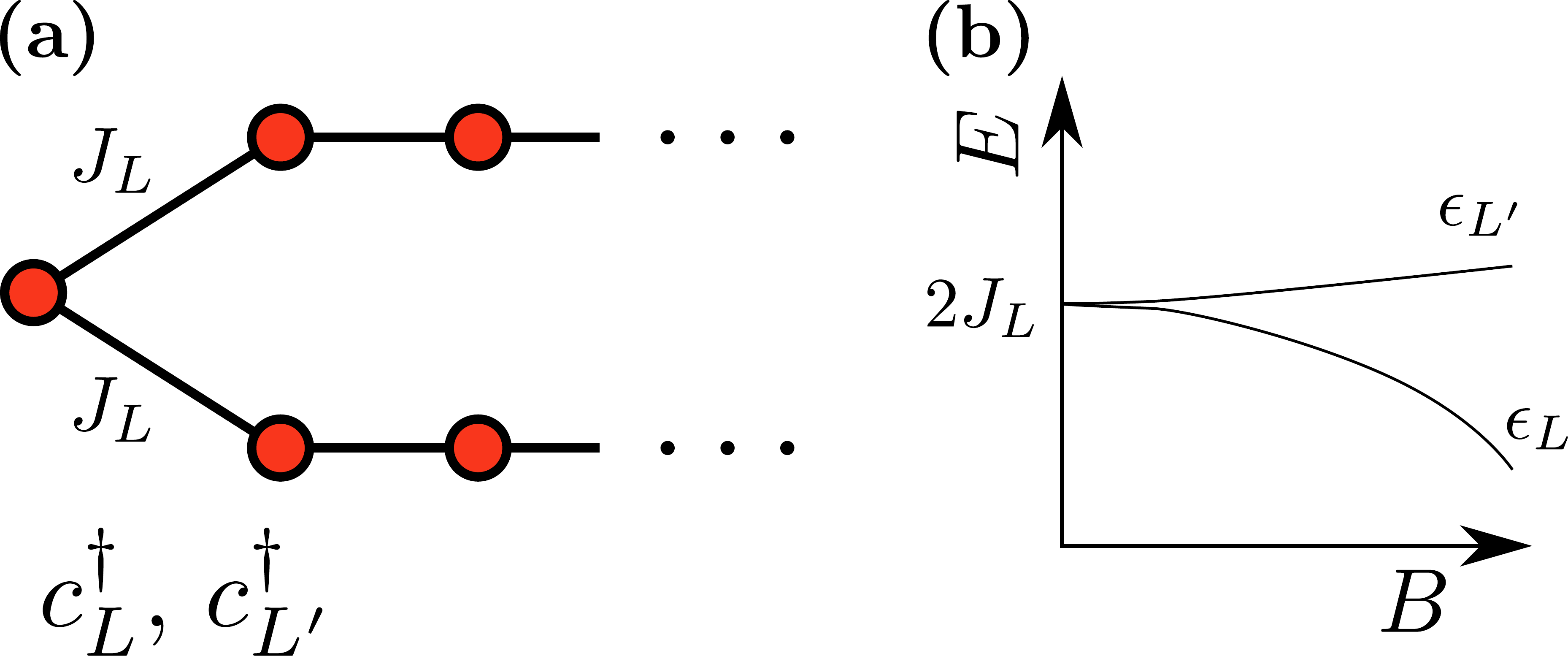}
        \caption{The bonds $J_L$ support a pair of fermion bound states $c^\dagger_L$ and $c^\dagger_{L'}$, shown in {\bf (a)} with energies $\epsilon_L(B)$ and $\epsilon_{L'}(B)$ shown in {\bf (b)}.}
        \label{fig:bdgl}
\end{figure}

\begin{figure} 
        \includegraphics[width=\columnwidth]{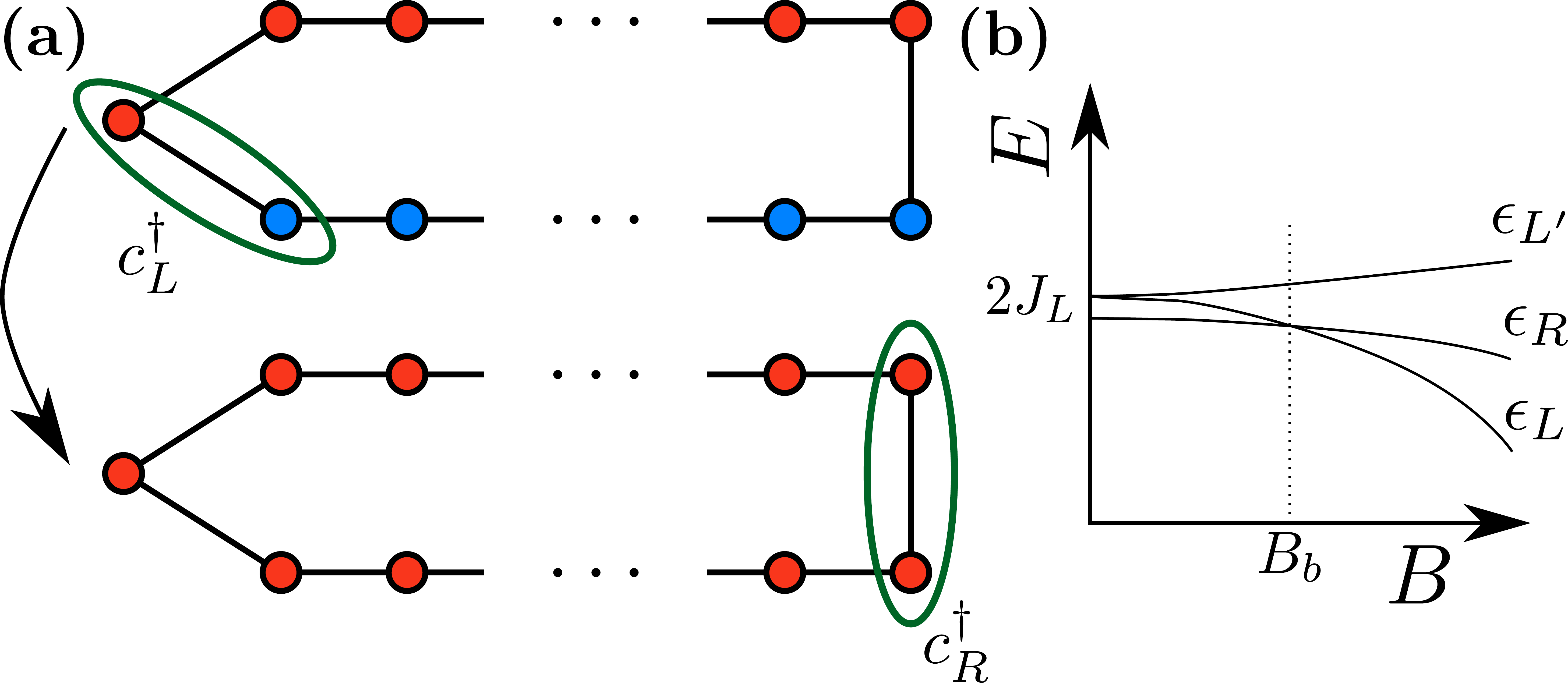}
        \caption{{\bf  The BdG-boundstate crossing} {\bf (a)} A fermion bound state (a dressed 0-dimensional domain wall) tunnels across the graph from left to right. It flips all of the qubits along its way. {\bf (b)} Bound-state energies as a function of $B$. As external field $B$ is lowered, the defect $\hat{c}_R^\dag$ becomes energetically favorable at $B_b$ and the tunneling shown in panel {\bf (a)} takes place. }
        \label{fig:crossingspectrum}
\end{figure}

The origin of the spin glass bottleneck becomes clear once one realizes that the doubling of the $J_L$-bond means that we have left-out a second $L$-localized boundstate $\hat{c}_{L'}$, which deflects the energy dispersion curve $\epsilon_L(B)$ of the $\hat{c}_L$-defect downwards (see Figure \ref{fig:bdgl}), so that, for $JJ_R > J_L^2$, it crosses with the energy of the $\hat{c}_R$ defect 
at a tunable $B\equiv B_b$ value of the transverse-field within the spin-glass phase of the anneal:
\begin{equation}
B_b \equiv \frac{1}{J_R} \frac{(J^2-J_L^2)(J_L^2-J_R^2)}{J^2+J_R^2-2J_L^2}<B_c \ .\label{eq:bottleneck_point}
\end{equation}
A rigorous derivation of the identity in Eq. \eqref{eq:bottleneck_point} is carried out in Appendix \ref{app:low}. The level crossing produces the situation in the odd-fermion sector depicted in Figure \ref{fig:crossingspectrum}, where a forced excitation (forced by parity constraints) must quantum-mechanically tunnel from left to right in the graph. A similar situation occurs in the even-fermion sector, leading to a second level crossing at $B\equiv B_b$. So, in total, at $B=B_b$, two pairs of energy levels cross.\\

\subsubsection{Scaling of the gap}
The previous analysis was only exact in the limit $\mN \to\infty$. At finite $\mN$, in each parity sector, there will be hybridization between boundstates localized at opposite ends of the graph (due to finite-size effects), and so, directly at the minimum gap region, the true fermonic eigenmodes will sweep rapidly through a mixture of left- and right-modes, producing an avoided (Landau- Zener) crossing. For example, in the even-fermion sector, one can define a Landau-Zener approach angle $$\tan 2\theta_{LZ}(B)\equiv \frac{\Delta(B)}{\epsilon_{R}^+(B)-\epsilon_{L}^+(B)} \ , $$
where $\Delta(B)$ is defined as the (exponentially-small) overlap between the left- and right-localized even-partity fermion boundstates. In terms of $\theta_{LZ}$, the two lowest-energy excitations become:
\begin{align}
\hat{c}_{0,+}^\dag&=\hat{c}_{R,+}^\dag\sin\theta_{LZ}   + \hat{c}_{L,+}^\dag\cos\theta_{LZ} \ , \\
\hat{c}_{1,+}^\dag&=\hat{c}_{L,+}^\dag\sin\theta_{LZ}- \hat{c}_{R,+}^\dag\cos\theta_{LZ} \ .
\end{align}
In contrast, the odd-parity bound states do not hybridize, as they have different parities under spatial reflection $j\to \mN-(j-1)$ (see
Appendix~\ref{app:low}), and the crossing there is exact, for all $\mN<\infty$. Therefore, at $B=B_b$, in the large-$\mN$ limit, there is a pair of crossings, with each crossing occurring
in a distinct eigenspace of $(-1)^{N_F}$, i.e. each crossing is labelled by a distinct $\mathbb{Z}_2$ quantum number, see bottom portion of Fig.~\ref{fig:lowenergy}. The scaling of the hybridization of the even-parity boundstates at the bottleneck location gives the inverse
QA runtime
\begin{equation}
\Delta(B_b) \underset{\mN\to \infty}{\propto} \mathcal O\left(\frac{J(J_L^2-J_R^2)}{J_R(J^2-J_L^2)} \right)^\mN \ ,\label{eq:rigorous_gap_scaling}
\end{equation}
which is exponentially small in $\mN$, as desired. Note that the spectral data in the odd-parity sector is irrelevant in analyzing the time-complexity of the closed-system annealing dynamics, as the odd-parity sector is never visited during the coherent evolution. This is because the initial state of the QA protocol always has even spin-flip parity (see, e.g. \cite{dziarmaga_dynamics_2005}), and spin-flip parity is conserved throughout the annealing schedule.

\section{Analysis of flux qubit noise at the annealing bottleneck} \label{sec:noise}

In this section we turn our attention to the effects of ambient flux qubit
noise at a frustrated spin glass bottleneck. In the D-Wave 2X quantum annealer, fluctuations in onsite qubit flux bias form the dominant source of noise \cite{boixo_computational_2016,
amin_role_2009}. This flux bias noise is accurately modelled by the following system-bath Hamiltonian:\\
\begin{eqnarray}
\hat{H}_\text{D-Wave} &\equiv& \hat{H}_0+\sum_{j=1}^\mN\hat{Q}_j\hat{\sigma}^z_j + \hat{H}_B \ , \label{eq:dwave_ham1}\\
\hat{Q}_j &\equiv& \sum_u \lambda_u(\hat{b}_{j,u}+\hat{b}_{j,u}^\dag) \ , \label{eq:dwave_ham2}
\\ \hat{H}_B &=& \sum_{j,u}\hbar \omega_u (\hat{b}_{j,u}^\dag \hat{b}_{j,u}+1/2) \ ,\label{eq:dwave_ham3}
\end{eqnarray}
where the bath operators $\{\hat{b}_{j,u}\}_u$ for each flux
qubit satisfy standard bosonic commutation relations
$[\hat{b}_{j,u},\hat{b}_{j,u'}^\dag]=\delta_{u,u'}$, and $\omega_u$ is the frequency of mode
$u$ and $\lambda_u$ sets the interaction strength between that mode and its corresponding qubit. For weak noise strength, the effects of the flux bias fluctuations on the system dynamics are uniquely characterized by the noise spectral density of these fluctuations:
\begin{align}
    S(\omega)\equiv \int_0^\infty dt e^{i\omega t}\langle e^{iH_Bt/\hbar}\hat{Q}_je^{-iH_Bt/\hbar}\hat{Q}_j\rangle \ ,
\end{align}
which we assume to be identical for each qubit. For simplicity, in this study we assume the spectral density of the noise to be {\it Ohmic}:
\begin{equation}
S(\omega) \equiv \hbar^2\frac{\eta \omega
e^{-\omega\tau_c}}{1-\exp(-\hbar\omega/k_BT)}\label{eq:spectral_density}
\end{equation}
at temperature $T$. Here, $\eta$ characterizes the strength of the fluctuations in the flux bias, and is typically measured in macroscopic resonant tunneling experiments on the individual qubits in the processing unit. For the D-Wave 2X annealer at NASA QuAIL, $\eta$ was measured to be $\sim 0.24$ in the regime $B\equiv 0$ \cite{boixo_computational_2016}. We use this value of $\eta$ for our numerical simulations of the D-Wave 2X at Los Alamos National Laboratory. \begin{figure}
        \includegraphics[width=\columnwidth]{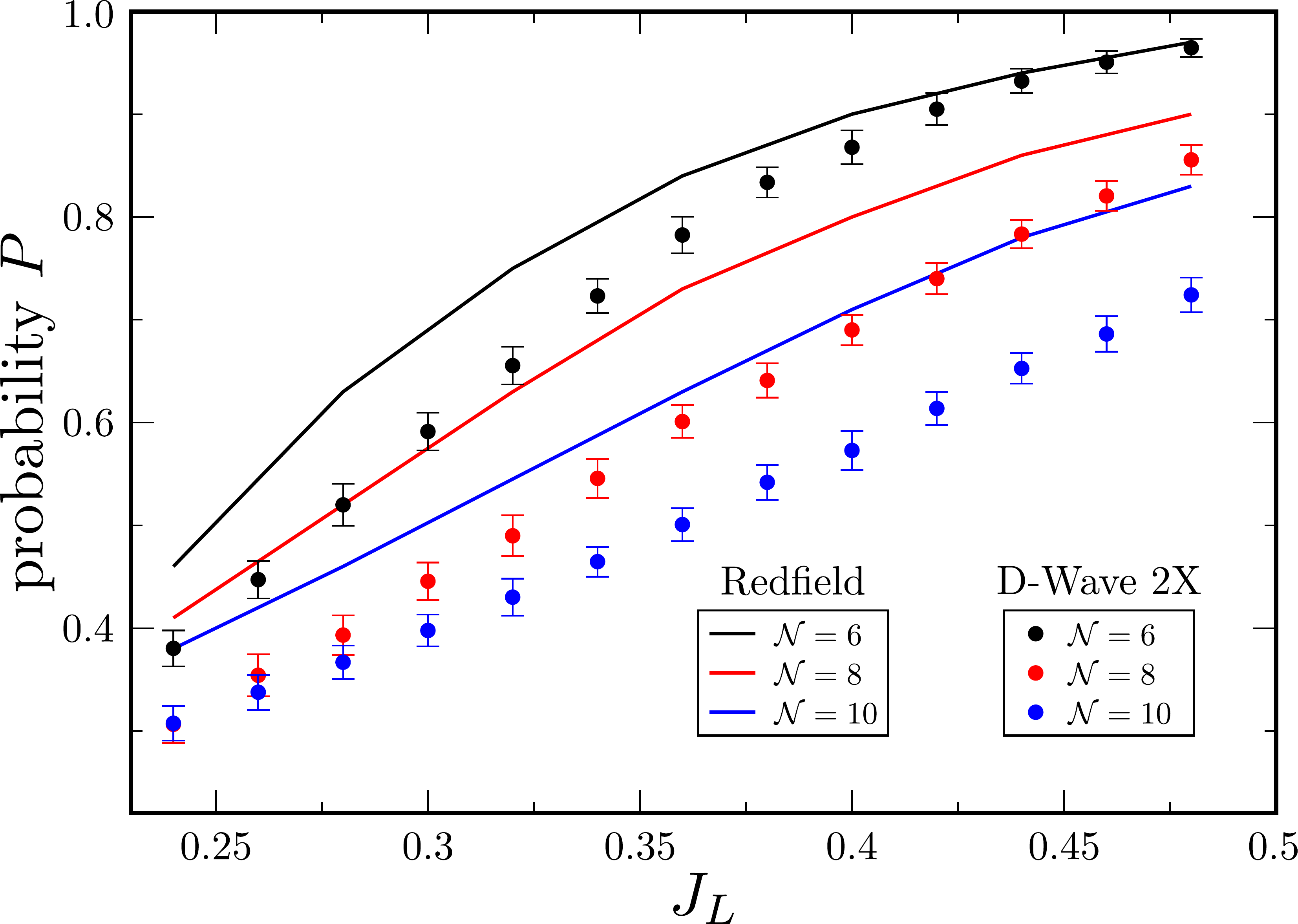}
        \caption{Probability of success vs. $J_L$ for the D-Wave 2X. The plot shows probability $P$ that the D-Wave 2X at Los Alamos National Laboratory finds the optimal answer to the Frustrated Ring MAXCUT problem, as a function of the parameter $J_L$ of the Frustrated Ring for system size $\mN = 6,8,10$. We considered the Frustrated Ring defined by the following coupling parameters: $J=1$, $J_R=0.2$ and $0.2<J_L<0.5$. The simulation on D-Wave 2X was run with annealing time $t_\text{QA} = 5~\mu$s and temperature $T=15.5$ mK.  We find qualitative agreement of numerical Redfield simulations with the D-Wave 2X.}
        \label{fig:bottleneck}
\end{figure}

We treat the open-system dynamics specified by Eqs. (\ref{eq:dwave_ham1}-\ref{eq:dwave_ham3}) in the Bloch-Redfield
approximation, which, at the qualitative level, closely predicts the probability $P$ that the D-Wave 2X
machine returns a global minimum of the Frustrated Ring MAXCUT problem:
\begin{align}
    P&\equiv \langle 00\cdots 0|\hat{\rho}_f|00\cdots 0\rangle+\langle 11\cdots 1|\hat{\rho}_f|11\cdots 1\rangle,\label{eq:success_0}
\end{align}
where here, $\hat{\rho}_f$ is the collective density matrix of the qubits in the annealer at the end of the anneal, and $\{ 000\cdots 0,111\cdots 1\}$ is the complete set of bit strings which solve the Frustrated Ring MAXCUT problem. The actual success probability observed in the D-Wave 2X quantum annealer, compared with the simulated success probability (according to the Bloch-Redfield simulation of the quantum processor) is plotted in Fig. \ref{fig:bottleneck}. \\

Assuming the annealing processor is completely incoherent, i.e. the off-diagonal matrix elements of $\hat{\rho}$ vanish in the energy eigenbasis, then the Redfield master equation degenerates into a kinetic equation involving tunneling rates between
instantaneous eigenstates. These are given by Fermi's golden rule as (c.f. \cite{amin_role_2009}):
\begin{eqnarray}
\Gamma_{i\to f} &=& \frac{1}{2\hbar^2}S(\omega_{i\to f}) \cdot O_{i\to f} \ , \\
O_{i\to f} &\equiv& \sum_{j=1}^\mN |\bra{\Psi_i} \hat{\sigma}^z_j \ket{\Psi_f}|^2 \ ,
\end{eqnarray}
where here, $\omega_{i\to
f}$ is the gap frequency $(E_i-E_f)/\hbar$ between the initial $\ket{\Psi_i}$
and final eigenstate $\ket{\Psi_f}$. Directly at the bottleneck in our spin glass benchmark, we will find this incoherent evolution to be analytically solvable in the large-$\mN$ limit. Symmetry simplifies the problem: since $\hat{\sigma}^z$ flips the
$\mathbb{Z}_2$ quantum number corresponding to spin-flip parity, the
relevant form-factors that need to be calculated are (assuming the temperature is
sufficiently low so that we can assume that the four-lowest levels are populated during the evolution):
\begin{eqnarray}
O_{0^+\to L^-} &=& \sum_j |\bra{\Psi_0^+} \hat{\sigma}^z_j \ket{\Psi_L^-} |^2 \ ,\label{eq:tunneling1}\\
O_{0^+\to R^-} &=& \sum_j |\bra{\Psi_0^+} \hat{\sigma}^z_j \ket{\Psi_R^-} |^2 \ \label{eq:tunneling2},\\
O_{1^+\to L^-} &=& \sum_j |\bra{\Psi_1^+} \hat{\sigma}^z_j \ket{\Psi_L^-} |^2 \ , \label{eq:tunneling3}\\
O_{1^+\to R^-} &=& \sum_j |\bra{\Psi_1^+} \hat{\sigma}^z_j \ket{\Psi_R^-} |^2 \ \label{eq:tunneling4}.
\end{eqnarray}
All other matrix elements vanish by symmetry, because $\hat{\sigma}^z$ mixes
fermionic $\mathbb{Z}_2$ parity symmetry. To calculate the transition rates,
we expand them so that they are written completely in terms of the basis
$\{\ket{\Psi_L^-},\ket{\Psi_R^-}, \ket{\Psi_L^+}, \ket{\Psi_R^+}\}$ of
crossing states:
\begin{equation}
\begin{split}
O_{0^+\to {L,R}^-} = \sum_j & \left|  \sin\theta_{LZ} \bra{\Psi_R^+} \hat{\sigma}^z_j \ket{\Psi_{L,R}^-} \right.  \\
 & + \left. \cos\theta_{LZ} \bra{\Psi_L^+} \hat{\sigma}^z_j \ket{\Psi_{L,R}^-} \right|^2 \ ,
\end{split}\label{eq:O0R}
\end{equation}

\begin{equation}
\begin{split}
O_{1^+\to {L,R}^-} = \sum_j & \left| \cos\theta_{LZ} \bra{\Psi_R^+} \hat{\sigma}^z_j \ket{\Psi_{L,R}^-}  \right. \\
 & \left. - \sin\theta_{LZ} \bra{\Psi_L^+} \hat{\sigma}^z_j \ket{\Psi_{L,R}^-} \right|^2 \ .
\end{split}\label{eq:O1R}
\end{equation}
Some comments are in order. In general, tunneling due to non-fermionizable
(i.e. $\hat{\sigma}^z$) noise in a one-dimensional quantum spin glass is analytically
intractable due to the fact that the interacting portion of the Hamiltonian in Eq. \eqref{eq:dwave_ham1} maps to a Jordan-Wigner string
\begin{equation}
\sum_{j}\hat{\sigma}^z_j \hat{Q}_j = \sum_{j}(-i)^j \hat{\gamma}_1 \cdots \hat{\gamma}_{2j-1}
\sum_u\lambda_u(\hat{b}_{j,u}+\hat{b}_{j,u}^\dag) \ .
\end{equation}
Therefore, solving the full D-Wave dynamics even at the perturbative,
Markovian level is widely considered to be analytically intractable
\cite{keck_dissipation_2017, deng_decoherence_2013}. For example, one of the
tunneling matrix elements above involves an inner-product of the form
\begin{equation}
\bra{\Psi_R^+} \hat{\sigma}^z_j \ket{\Psi_{R}^-} = \bra{\Omega_+} (-i)^j\hat{\gamma}_1\cdots \hat{\gamma}_{2j-1} \hat{c}_{R,-}^\dag \ket{\Omega_-} \ .\label{eq:interaction_ham}
\end{equation}
The Jordan-Wigner string in the above equation means that the matrix element evaluates to a determinant
of growing size. Furthermore, the parity-dependent boundary conditions
and broken translational invariance of the model, make the attempts of
obtaining the closed-form solution futile. What is more, the perturbative
treatment of this matrix element is also ill-fated because of the short
radius of convergence of perturbation theory in $B$ in generic spin glasses
(inevitably occurring at the first closing of the gap). In this work, we treat
the transverse-field at the non-perturbative level by performing a field-theoretic calculation (see Appendix \ref{app:calc}), leading to an analytical
understanding of flux qubit noise at a spin glass annealing bottleneck.

\subsection{Non-perturbative large-$\mN$ calculation of tunneling rates}
We begin by computing the
off-diagonal, i.e. $\ket{\Psi_L}\to
\ket{\Psi_R}$, matrix elements. These vanish in the large-$\mN$
limit as a consequence of Lieb-Robinson bounds \cite{bravyi_topological_2010}. For example, we can factor
\begin{equation}
\bra{\Psi_L^+}    \hat{\sigma}_j^z    \ket{\Psi_R^-} =
\bra{\Psi_L^+(0)} \hat{\sigma}_j^z(B) \ket{\Psi_R^-(0)} \ ,
\end{equation}
where $\hat{\sigma}_j^z(B) = \hat{U}^\dag(B)\hat{\sigma}^z_j \hat{U}(B)$, and the unitary $\hat{U}(B)$ has the general form
\begin{align}
    \hat{U}(B)&\equiv \mathcal T[e^{-i\int_0^B dB' \widetilde{H}(B')}]
\end{align}
with $\widetilde{H}$ defined in \cite{bravyi_topological_2010}. Now, suppose $B<B_b$ is fixed. Since $\hat{H}$ has a spectral gap which is at least $\mathcal O(1)$ for all $B'<B$, $\hat{U}(B)$ is a constant-depth unitary
circuit \cite{chen_local_2010}. That is, $\hat{\sigma}^z_j(B)$, up to exponentially small corrections
constant in $\mN$, is supported on a region of constant size. However,
$\ket{\Psi_L^+}$ and $\ket{\Psi_R^-}$ at $B=0$ are
separated by $\sim \mN/2$ spin flips. Therefore, for any fixed $B<B_b$, the matrix elements mixing $\ket{\Psi_L}$
with $\ket{\Psi_R}$ are exponentially small in $\mN$, i.e.
\begin{align}
    \bra{\Psi_R^+} \hat{\sigma}^z_j \ket{\Psi_L^-}&\underset{\mN\to \infty}{\sim} \mathcal O(e^{-c\mN/2}) \ ,\label{eq:off_diag_elts1}
\end{align}
with $c$ a constant. Within a sufficiently small neighborhood of the crossing, we can assume the diabatic crossing states $\ket{\Psi_L^\pm}$
and $\ket{\Psi_R^\pm}$ to be approximately independent of $B$ (with the $B$-dependence of the true eigenstates due to mixing within the subspace spanned by these crossing states). Therefore, for $B>B_b$, sufficiently near the crossing, the off-diagonal matrix elements are also exponentially small. Via the exact same reasoning, we also have the asymptotic behavior
\begin{align}
    \bra{\Psi_R^-} \hat{\sigma}^z_j \ket{\Psi_L^+}&\underset{\mN\to \infty}{\sim} \mathcal O(e^{-c'\mN/2}) \ ,\label{eq:off_diag_elts2}
\end{align}
for the other off-diagonal matrix element with $c'$ another constant. Again, this follows from the fact that at $B=0$ the state $|\Psi_R^-\rangle$ is separated from the state $|\Psi_L^+\rangle$ by $\sim \mN/2$ spin flips.

\begin{figure} 
        \includegraphics[width=\columnwidth]{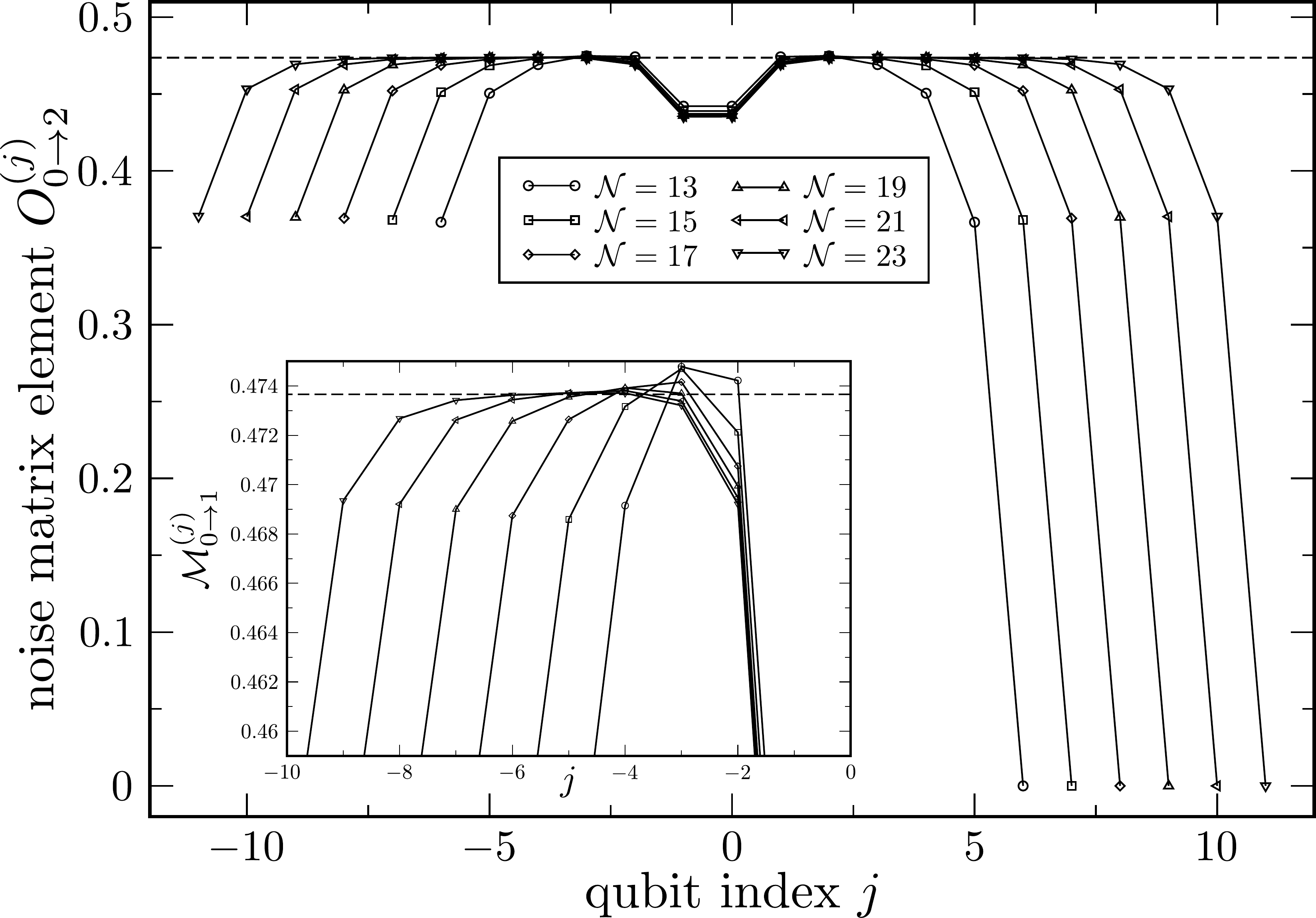}
        \caption{ Noise amplification at a spin-glass bottleneck. Very close to the annealing bottleneck, provided that the qubit index $j$ is in the bulk of the chain, the noise matrix element $O_{0\to 2}^{(j)}=\theta(B_b-B) |\bra{\Psi_0^+} \hat{\sigma}^z_j \ket{\Psi_L^{-}}|^2+\theta(B-B_b)|\bra{\Psi_0^+} \hat{\sigma}^z_j \ket{\Psi_R^{-}}|^2$ (c.f. Eqs. (\ref{eq:tunneling1}-\ref{eq:tunneling4})) approaches the predicted value of $(1 - (B_b/J)^2)^{1/4}/2$ (dashed line). This is $1/2$ the squared magnetization of the uniform quantum Ising chain. Therefore, the corresponding transition rate (which is a sum over all qubits) is $\mO(\mN)$. See text for details.}
        \label{fig:gapped}
\end{figure}

In summary, the off-diagonal contributions to multiqubit tunneling asymptotically vanish near the crossing in the large-$\mN$ limit. Therefore, provided that we are sufficiently close to the crossing point so that neither $\tan\theta_{LZ}$ nor $\cot\theta_{LZ}$ are exponentially small in $\mN$, we can ignore these off-diagonal contributions. In this limit, the large-$\mN$ tunneling form factors have the asymptotics
\begin{align}
    O_{0^+\to R^-}&\underset{B\to B_b}{\sim} \sum_{j} \sin^2\theta_{LZ}|\langle \Psi_R^+|\sigma^z_j|\Psi_R^-\rangle|^2 \ ,\label{eq:form_factors_near_the_crossing1} \\
    O_{0^+\to L^-}&\underset{B\to B_b}{\sim} \sum_{j}\cos^2\theta_{LZ} |\langle \Psi_L^+|\sigma^z_j|\Psi_L^-\rangle|^2 \ .\label{eq:form_factors_near_the_crossing2}
\end{align}
with analogous expressions for $O_{1^+\to R^-,L^-}$. To calculate the simplified form factors given by Eqs. (\ref{eq:form_factors_near_the_crossing1}-\ref{eq:form_factors_near_the_crossing2}), we begin with the following generic observation: for each time-dependent annealing Hamiltonian $H_0(t)$ specified in Eq. (\ref{eq:QA_ham}), let $\widetilde{H}_0(t)$ denote the annealing Hamiltonian obtained by flipping the sign of $J_\mN$. Upon taking the Jordan-Wigner transformation (c.f. Eq. (\ref{eq:HFmodel})), we then have the
following relations:
\begin{equation} \label{eq:intertwining}
H^\pm_0(t)=\widetilde{H}^{\mp}_0(t) \ . 
\end{equation}
These relations are completely general and hold for any quantum spin glass on a
2-regular graph. Crucially, if $H_0$ is frustrated, $\widetilde{H}_0$ lacks frustration, and thus perhaps easier to characterize. For the Frustrated Ring, we can use the above relation to obtain a complete solution to
the incoherent tunneling rates (c.f. Eqs. (\ref{eq:tunneling1}-\ref{eq:tunneling4})) in the large-$\mN$ limit.

Indeed, in Appendix \ref{app:calc}, we derive the following crucial identities relating low-lying eigenstates of the Frustrated Ring spin glass benchmark with those of its ferromagnetic counterpart $\widetilde{H}$:
\begin{align}
    \langle \Psi_{R,L}^+|\sigma^z_j|\Psi_{R,L}^-\rangle &= -\langle \widetilde{\Psi}_{R,L}^-|\sigma^z_j|\widetilde{\Psi}_{R,L}^+\rangle + \mathcal O(e^{-\kappa_R |j-j_R|}) \ ,\label{eq:intertwining_the_states}
\end{align}
here, $j_R$ is the position of the antiferromagnetic coupler $J_R$, and $\kappa_R$ is the wavenumber of the boundstate $\hat{c}_R^\dag$. Note that these errors are localized at the position of the $J_R$-coupler, and thus do not grow if we sum over all qubits in the graph.

To summarize, by the replacements $\ket{\Psi_{L,R}^\pm} \to
\ket{\widetilde{\Psi}_{L,R}^\pm}$, we can relate our frustrated tunneling
form factors to those in an unfrustrated spin system $\widetilde{H}$ (c.f. Eq. \ref{eq:intertwining}),
at the cost of inducing an error which does not grow with the total number
of qubits $\mN$. We thus have
\begin{align}
    O_{0^+\to R^-}&\underset{\mN\to \infty}{\sim} \sum_{j} \sin^2\theta_{LZ}|\langle \widetilde{\Psi}_R^+|\sigma^z_j|\widetilde{\Psi}_R^-\rangle|^2,\label{eq:no_frustration1} \\
    O_{0^+\to L^-}&\underset{\mN\to \infty}{\sim} \sum_{j}\cos^2\theta_{LZ} |\langle\widetilde{ \Psi}_L^+|\sigma^z_j|\widetilde{\Psi}_L^-\rangle|^2,\label{eq:no_frustration2}
\end{align}
with analogous expressions for $O_{1^+\to R^-,L^-}$. Note that the above asymptotics are not valid unless the expressions in Eqs. (\ref{eq:no_frustration1}-\ref{eq:no_frustration2}) are asymptotically greater than $\mathcal O(1)$; we will find that this is the case (c.f. Eqs. (\ref{eq:culm1}-\ref{eq:culm2})), so that our calculation is self-consistent. The frustration-free version ($\widetilde{H}_0$) of our problem is much easier to solve: in particular, any frustration-free spin system is gauge-equivalent to a ferromagnet via a local $\mathbb{Z}_2$ gauge-transformation of the form
\begin{align}
    \hat{U}(g)\equiv \prod_{j}(\sigma^x_j)^{g_j},
\end{align}
with $g_j\in \{0,1\}$. In fact, we find that the matrix elements in Eqs. (\ref{eq:no_frustration1}-\ref{eq:no_frustration2}) are related to the {\it spontaneous magnetization} of the ferromagnetic spin chain $\widetilde{H}$. Indeed, at $B=0$,
we have
\begin{eqnarray}
\bra{\widetilde{\Psi}_R^+} \hat{\sigma}^z_j \ket{\widetilde{\Psi}_R^-} &=& 1 \ , \label{eq:zerofield1}\\
\bra{\widetilde{\Psi}_L^+} \hat{\sigma}^z_j \ket{\widetilde{\Psi}_L^-} &\sim& \text{sgn}(j) \ .
\label{eq:zerofield2}
\end{eqnarray}
Therefore, at zero transverse field, the above
matrix elements have the physical meaning of being the local magnetization
of each state (in Eqs. (\ref{eq:zerofield1}-\ref{eq:zerofield2}), we have implicitly re-indexed the qubits from $j=-n,\cdots,0,\cdots, n$, where $\mN=2n+1$). Exact analytical expressions for the matrix elements in Eqs. (\ref{eq:zerofield1}-\ref{eq:zerofield2}) in the more general case $B>0$ can be obtained by performing a field-theoretic calculation in the corresponding two-dimensional classical Ising model (see Appendix~\ref{app:calc}). There, we find that the absolute values of the
magnetizations in Eqs. (\ref{eq:zerofield1}-\ref{eq:zerofield2}) approach the bulk value 
$M\equiv (1-(B/J)^2)^{1/8}$. Note that $M$ is equal to the bulk spontaneous magnetization of a quantum Ising chain with uniform ferromagnetic coupling $J$ \cite{iorgov_form_2011}. A transfer matrix argument in Appendix \ref{app:calc} is used to show that this convergence is exponentially fast in the distance from the $J_L,J_R$ defects. We confirm this prediction with exact diagonalization  for $\mN$ up to $23$ sites, see Fig. \ref{fig:gapped}.

\begin{figure} 
        \includegraphics[width=\columnwidth]{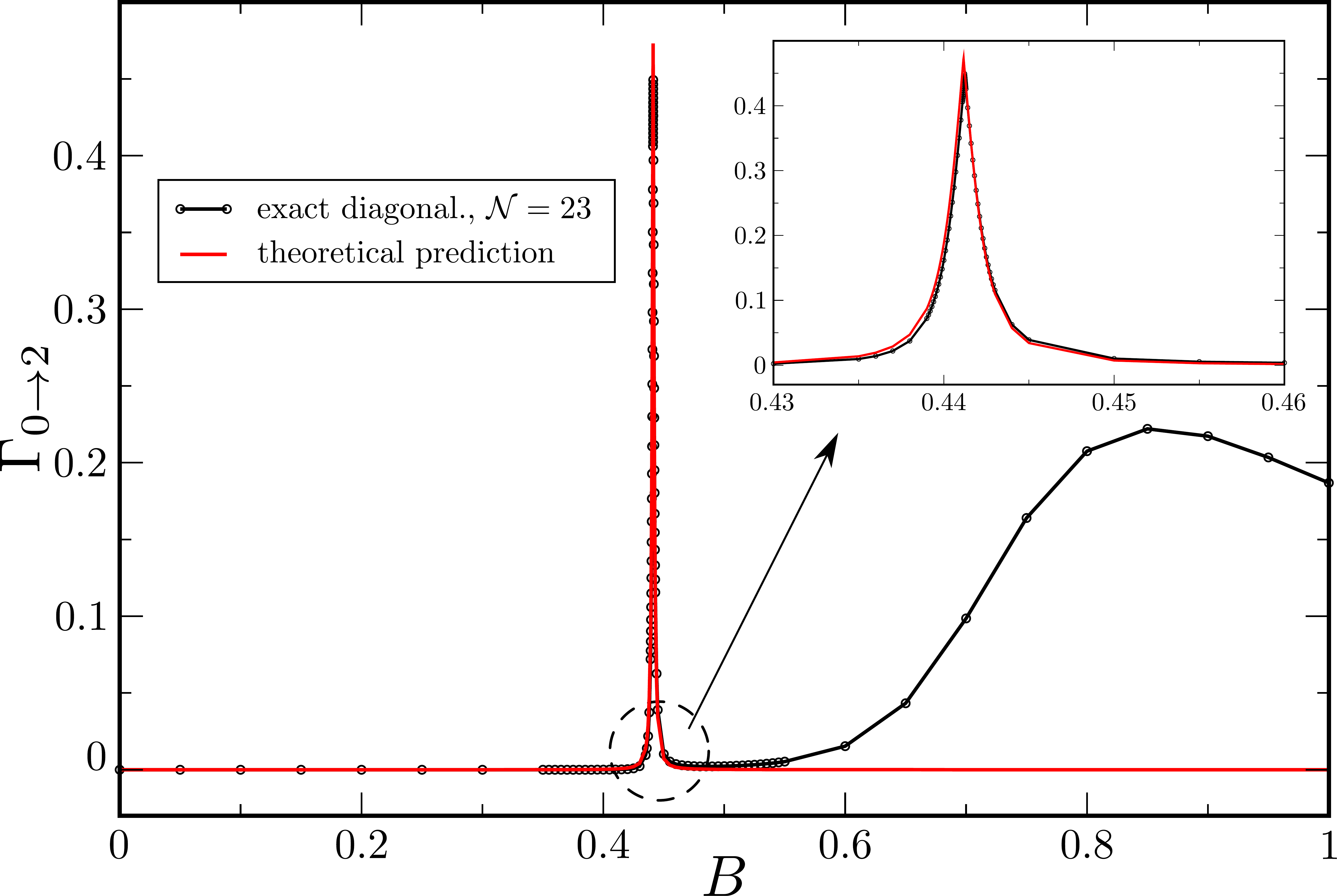}
        \caption{Analytical description of a spin glass bottleneck of quantum annealing. Very close to the bottleneck, the multiqubit tunneling rate $\Gamma_{0\to 2}=\theta(B_b-B)\Gamma_{0^+\to L^-}+\theta(B-B_b)\Gamma_{0^+\to R^-}$ (black) is given exactly by the analytical expression (\ref{eq:culm1}-\ref{eq:culm2}) (red). We utilize this analytical understanding to probe the dynamics of a finite-temperature anneal through this region.}
        \label{fig:qabott}
\end{figure}

As a result, the multiqubit tunneling rates at the spin glass bottleneck have the following large-$\mN$ asymptotic form:
\begin{eqnarray}
\Gamma_{0^+\to L^-} &\underset{\mN\to \infty}{\sim}& S(\omega_{0^+\to L^-})\frac{\mN M^2}{2\hbar^2}\cos^2\theta_{LZ} \label{eq:culm1} \ , \\
\Gamma_{0^+\to R^-} &\underset{\mN\to \infty}{\sim}& S(\omega_{0^+\to R^-})\frac{\mN M^2}{2\hbar^2}\sin^2\theta_{LZ} \label{eq:culm2} \ , \end{eqnarray}
which is $\mathcal O(\mN)$, with similar expressions for $\Gamma_{1^+\to L^-,R^-}$. Again,
$M\equiv (1-(B_b/J)^2)^{1/8}$, the bulk value of the spontaneous
magnetization of the one-dimensional transverse-field Ising model, sets the coefficent of scaling. The above expressions are in excellent agreement
with exact-diagonalization for $\mN$ up to $23$ sites, as shown in
Fig.~\ref{fig:qabott}.

The physical implications
of Eqs. (\ref{eq:culm1}-\ref{eq:culm2}) above are clear: in the Landau-Zener formalism, at
$t=-\infty$, $(\sin\theta,\cos\theta)=(0,1)$, and then, at $t=\infty$, we have that
$(\sin\theta,\cos\theta)=(1,0)$. Therefore, at $t=\pm \infty$, the transition rate
is predominantly the exponentially-small $\ket{\Psi_L} \to
\ket{\Psi_R}$ cross-terms (which were neglected in deriving (\ref{eq:culm1}-\ref{eq:culm2})), and the approximation breaks down. However,
at the bottleneck $t\sim 0$,  the transition rate quickly reaches a peak
which is asymptotic to $\mN M^2/2$, times the noise-spectral density evaluated at the minimum gap frequency.

\section{Large-$\mN$ limit of quantum annealing through the bottleneck at finite-temperature} \label{sec:largeN}
The simple analytical formulae (\ref{eq:culm1}-\ref{eq:culm2}) demonstrate the existence of a linear $\mathcal O(\mN)$ tunneling peak near a quantum annealing bottleneck, and establish the relation of the scaling coefficient to a suitably defined bulk spontaneous magnetization, $M$, in our spin glass benchmark. Using these asymptotics, we can rigorously analyze the effects of flux-bias noise on
a quantum annealing chip at a spin glass bottleneck, in the
limit that the number of qubits tends to infinity.

Assuming that the annealing processor is fully incoherent, the density matrix $\rho$ of the system is diagonal in the eigenbasis $\{\ket{\Psi_0^+}, \ket{\Psi_1^+}, \ket{\Psi_R^-}, \ket{\Psi_L^-} \}$ of the coherent portion $\hat{H}_0$ of the quantum annealing Hamiltonian (\ref{eq:dwave_ham1}):
\begin{equation}
\begin{split}
\hat{\rho} &= P_0^+ \ket{\Psi_0^+}\bra{\Psi_0^+} + P_1^+ \ket{\Psi_1^+}\bra{\Psi_1^+} \\
&+ P_R^- \ket{\Psi_R^-}\bra{\Psi_R^-} + P_L^-
\ket{\Psi_L^-}\bra{\Psi_L^-} \ .
\end{split} 
\end{equation}
Under those assumptions, the density matrix satisfies incoherent time-evolution
in terms of rates of the form $\Gamma_{i\to j}$
\begin{equation}
\begin{split}
\partial_t P_0^+ =&\Gamma_{L^-\to 0^+}P_L^-+\Gamma_{R^-\to 0^+}P_R^- \\
&- \sum_{L^-,R^-}\Gamma_{0^+\to R^-,L^-}P_0^+ \ ,\label{eq:pme0}
\end{split}
\end{equation}
constituting a kinetic equation of Pauli type. Via Eqs. (\ref{eq:culm1}-\ref{eq:culm2}), the multiqubit tunneling rates in the above equation can then be computed exactly in the
large-$\mN$ limit, within a sufficient radius of the crossing point such that
the approximation (\ref{eq:form_factors_near_the_crossing1}-\ref{eq:form_factors_near_the_crossing2}) is valid. It is thus convenient
to define a {\it tunneling region} $t_i<t<t_f$ within which this assumption holds. In this case, the instantaneous gap is much smaller
than the temperature, and the Ohmic noise spectral density $S(\omega)$ defined in (\ref{eq:spectral_density}) saturates at its low-frequency value $
S(\omega)\underset{\hbar \omega/k_BT \to 0}{\equiv} S^{(0)}$. Therefore, in the tunneling region $t_i<t<t_f$, the incoherent master equation Eq. (\ref{eq:pme0}) takes the rather symmetric form
\begin{align}
&\frac{\partial P_0^+}{\partial t} \underset{\mN\to\infty}{\sim} \frac{S^{(0)}\mN M^2}{2\hbar^2}( P_L^-\cos^2\theta_{LZ}+P_R^-\sin^2\theta_{LZ}-P_0^+) \ ,\nonumber\\
&\frac{\partial P_1^+}{\partial t} \underset{\mN\to\infty}{\sim} \frac{S^{(0)}\mN M^2}{2\hbar^2}( P_L^-\sin^2\theta_{LZ}+P_R^-\cos^2\theta_{LZ}-P_1^+) \ ,\nonumber\\
&\frac{\partial P_R^-}{\partial t} \underset{\mN\to\infty}{\sim}\frac{S^{(0)}\mN M^2}{2\hbar^2}( P_0^+\sin^2\theta_{LZ}+P_1^+\cos^2\theta_{LZ}-P_R^-) \ ,\nonumber\\
&\frac{\partial P_L^-}{\partial t} \underset{\mN\to\infty}{\sim} \frac{S^{(0)}\mN M^2}{2\hbar^2}( P_0^+\cos^2\theta_{LZ}+P_1^+\sin^2\theta_{LZ}-P_L^-). \
\label{eq:pme}
\end{align}
The above master equation represents the open-system dynamics of a fully incoherent quantum annealing processor at the spin glass bottleneck, in the limit $\mN \to\infty$. We can vectorize the density matrix populations by defining $\mathbb{P}\equiv [P_0^+,P_1^+,P_R^-, P_L^-]^T$. The kinetic equation (\ref{eq:pme}) then takes the following matrix form:
\begin{align}
&\frac{\partial \mathbb{P}}{\partial t}
=
\frac{S^{(0)}\mN M^2}{2\hbar^2}
\left( \hat{\tau}^x_1 \left( \sin^2\theta_{LZ} +  \hat{\tau}^x_2\cos^2\theta_{LZ} \right)  - 1_{4\times 4}\right)\mathbb{P} \ .\label{eq:full}
\end{align}
where here, we have introduced a {\it pseudospin}
\begin{align}
    \hat{\tau}^x_1 &\equiv \begin{pmatrix}
    0&1\\1&0
    \end{pmatrix}\otimes 1_{2\times 2},\\
    \hat{\tau}^x_2 &\equiv 1_{2\times 2}  \otimes \begin{pmatrix}
    0&1\\1&0
    \end{pmatrix}.
\end{align}

\begin{figure} 
        \includegraphics[width=\columnwidth]{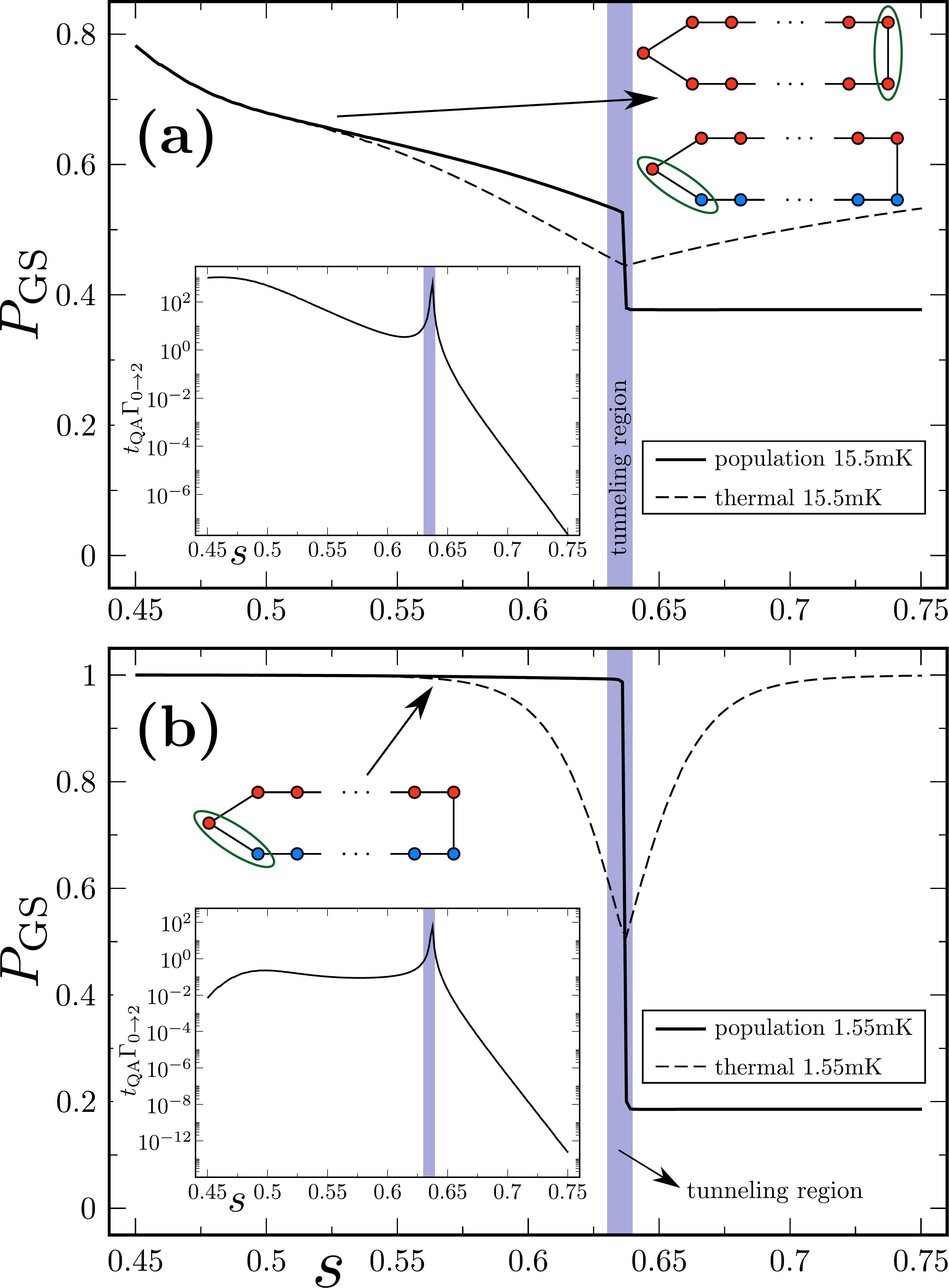}
        \caption{Freezing of annealing dynamics after a spin glass bottleneck. The plots show ground state population $P_{\textrm{GS}}\equiv P_0^++P_R^-$ as a function of dimensionless annealing parameter $s = t/t_\text{QA}$ for two temperatures  of the annealing chip, $T=15.5$ mK in panel {\bf (a)} and $T=1.55$ mK in panel {\bf (b)}. At typical operating temperatures of the D-Wave 2X, both $P_L$ and $P_R$ are nontrivial (c.f. Frustrated Ring depicted in panel {\bf (a)}) but at low-temperatures, $P_R$ is negligible (c.f. Frustrated Ring depicted in panel {\bf (b)}). In both regimes however, we note that nontrivial population transfer occurs directly at the bottleneck and then stops completely, making bottleneck physics particularly relevant to the performance of the quantum annealer. In the simulations, we considered an $\mN=8$ Frustrated Ring defined by $(J_R,J_L,J) = (0.2,0.24,1)$ and assumed annealing time $t_\text{QA} = 5~\mu$s. The insets show $t_\text{QA}\Gamma_{0 \to 2}$ (c.f. Eqs. (\ref{eq:culm1}-\ref{eq:culm2})) as a function of $s$.}
        \label{fig:hightemp}
\end{figure} 

\noindent The matrix representation Eq. (\ref{eq:full}) manifestly diagonalizes the Liouvillian for the effective classical master equation (\ref{eq:pme0}), allowing us to solve for the density matrix at all times. We begin by defining (anti-)symmetric combinations of the populations via 
\begin{align}
P_{\tau_1\tau_2} \equiv (P_0^++\tau_2P_1^+)+\tau_1(P_R^-+\tau_2P_L^-).
\end{align} 
Equivalently, $P_{\tau_1\tau_2}$ represents the projection of the vectorized density matrix $\mathbb{P}$ onto an arbitrary eigenspace of the time-dependent Liouvillian \eqref{eq:full}, with $\tau_1,\tau_2 \in \{+ 1,-1\}$ denoting the eigenvalues under application of $\tau_1^x,\tau_2^x$, respectively. In this eigenbasis, the ground
state population upon exiting the crossing is given by
\begin{equation}
P_{GS}(t_f)\equiv (P_R^- +P_0^+)\Bigr|_{t = t_f}
=\frac{1}{2}(1+P_{+-}(t_f)) \ .\label{eq:success_rate}
\end{equation}
The ground state population \eqref{eq:success_rate} upon exiting the crossing can be computed
analytically with arbitrary initial conditions. In particular, defining an effective rate $\Gamma_\text{eff}(t)\equiv S^{(0)}
M^2\cos^2\theta_{LZ}(B(t))/2\hbar^2$ which sets the timescale of the multiqubit dynamics, we then have 
\begin{align}
P_{GS}(t_f) & \underset{\mN\to\infty}{\sim}
\frac{1}{2}\left( 1 + e^{-2\mN \int_{t_i}^{t_f}\Gamma_\text{eff}(t)\,dt}P_{+-}(t_i) \right) \ .\label{eq:success}
\end{align}
Since this prediction only gives the ground-state population upon {\it exiting} the bottleneck region, the above analytical formula is relevant provided that
$P_{GS}(t=t_f)=P_{GS}(t=t_{QA})\equiv
P$ (c.f. Eq. (\ref{eq:success_0})). This is the case, e.g. if the density
matrix stops evolving non-trivially after the bottleneck (so-called
{\it freeze-out}, see e.g. \cite{marshall_thermalization_2017,amin_searching_2015,johnson_quantum_2011}). Crucially, we witness this stoppage of evolution in numerical
simulations of the Redfield master equation in the range $1.55-15.5$ mK, as
shown Fig.~\ref{fig:hightemp}. As we can see, up to $\mN=10$ (the performance
limit of our simulations), population transfer is nontrivial and limits the
performance of the quantum annealer. \\

We now focus on a specific set of initial conditions, supposing that the four
populations in the density matrix are {\it pairwise thermalized}. Before the
crossing point, we have $E_L^- - E_0^+\ll 1$  and $E_R^- - E_1 \ll 1$. Also,
the rates in the annealing process are larger earlier in the annealing
(due to larger transverse-field $B$). It is thus reasonable to assume that,
entering the crossing (i.e. at $t=t_i$), the density matrix populations satisfy
\begin{equation}
P_L^--P_0^+\ll 1,\quad
P_R^--P_1^+ \ll 1.
\end{equation}
Since the levels in each pair correspond to states with
the fermionic excitation localized at the same location, this is a reasonable
assumption to make. In this limit, $P_{+,-}(t_i)\to 0$, and so the success probability (\ref{eq:success}) is asymptotic to one half.

\section{Discussion} \label{sec:discussion}
In this paper we proposed and analytically solved a model that exhibits
the effects of frustration on bottlenecks of quantum annealing. By
investigating a simple class of one-dimensional annealing bottlenecks,
we are able to rigorously extract the scaling of tunneling rates caused by longitudinal qubit noise, in the large-$\mN$ (i.e. complexity-theoretic) limit. In conclusion, in our model, we have found that the effective noise spectral density
at an annealing bottleneck is $\mathcal O(\mN)$, which is of the order of the Hamming distance between the crossing
states. This exact result provides analytical confirmation of and is in agreement with results obtained in \cite{boixo_computational_2016, amin_role_2009}. Furthermore, in spite of the non-integrability of the flux bias noise (c.f. \eqref{eq:interaction_ham}), by treating the transverse-field $B$ at the non-perturbative level, we were able to extract the scaling coefficient accurate to all orders in the transverse-field and elucidate its relation to spontaneous magnetization.  Finally, we have confirmed using Redfield-type simulations that the ground-state occupation $P_{GS}$ upon exiting a spin glass bottleneck is
especially pertinent for the performance of a quantum annealer, and so we have identified a range of temperatures where our analytical scaling
formula should directly predict the annealing performance, in the limit of completely
incoherent system dynamics. In the future, we will investigate the effects of including coherences, i.e. off-diagonal elements of the density matrix, on the open-system dynamics. This will allow
us to gain more nuanced insight into the scaling performance of combinatorial
optimization and sampling problems on near-term quantum annealers.

\[\]

\begin{acknowledgments} This work was initiated at NASA QuAIL in 2016 under the USRA Quantum Academy program, and was also supported in part by the U.S. Department of Energy and in part by ASC Beyond Moore's Law project. LC was supported by the DOE through the J. Robert Oppenheimer fellowship. LC and AS also acknowledge support from the LDRD program at LANL. DR would like to thank Barry M. McCoy for a productive discussion. 
\end{acknowledgments}


\begin{thebibliography}{33}%
\makeatletter
\providecommand \@ifxundefined [1]{%
 \@ifx{#1\undefined}
}%
\providecommand \@ifnum [1]{%
 \ifnum #1\expandafter \@firstoftwo
 \else \expandafter \@secondoftwo
 \fi
}%
\providecommand \@ifx [1]{%
 \ifx #1\expandafter \@firstoftwo
 \else \expandafter \@secondoftwo
 \fi
}%
\providecommand \natexlab [1]{#1}%
\providecommand \enquote  [1]{``#1''}%
\providecommand \bibnamefont  [1]{#1}%
\providecommand \bibfnamefont [1]{#1}%
\providecommand \citenamefont [1]{#1}%
\providecommand \href@noop [0]{\@secondoftwo}%
\providecommand \href [0]{\begingroup \@sanitize@url \@href}%
\providecommand \@href[1]{\@@startlink{#1}\@@href}%
\providecommand \@@href[1]{\endgroup#1\@@endlink}%
\providecommand \@sanitize@url [0]{\catcode `\\12\catcode `\$12\catcode
  `\&12\catcode `\#12\catcode `\^12\catcode `\_12\catcode `\%12\relax}%
\providecommand \@@startlink[1]{}%
\providecommand \@@endlink[0]{}%
\providecommand \url  [0]{\begingroup\@sanitize@url \@url }%
\providecommand \@url [1]{\endgroup\@href {#1}{\urlprefix }}%
\providecommand \urlprefix  [0]{URL }%
\providecommand \Eprint [0]{\href }%
\providecommand \doibase [0]{http://dx.doi.org/}%
\providecommand \selectlanguage [0]{\@gobble}%
\providecommand \bibinfo  [0]{\@secondoftwo}%
\providecommand \bibfield  [0]{\@secondoftwo}%
\providecommand \translation [1]{[#1]}%
\providecommand \BibitemOpen [0]{}%
\providecommand \bibitemStop [0]{}%
\providecommand \bibitemNoStop [0]{.\EOS\space}%
\providecommand \EOS [0]{\spacefactor3000\relax}%
\providecommand \BibitemShut  [1]{\csname bibitem#1\endcsname}%
\let\auto@bib@innerbib\@empty
\bibitem [{\citenamefont {Boixo}\ \emph {et~al.}(2014)\citenamefont {Boixo},
  \citenamefont {Ronnow}, \citenamefont {Isakov}, \citenamefont {Wang},\ and\
  \citenamefont {Wecker}}]{boixo_evidence_2014}%
  \BibitemOpen
  \bibfield  {author} {\bibinfo {author} {\bibfnamefont {S.}~\bibnamefont
  {Boixo}}, \bibinfo {author} {\bibfnamefont {T.~F.}\ \bibnamefont {Ronnow}},
  \bibinfo {author} {\bibfnamefont {S.~V.}\ \bibnamefont {Isakov}}, \bibinfo
  {author} {\bibfnamefont {Z.}~\bibnamefont {Wang}}, \ and\ \bibinfo {author}
  {\bibfnamefont {D.~et.~al.}\ \bibnamefont {Wecker}},\ }\bibfield  {title}
  {\enquote {\bibinfo {title} {Evidence for quantum annealing with more than
  one hundred qubits},}\ }\href {\doibase 10.1038/nphys2900} {\bibfield
  {journal} {\bibinfo  {journal} {Nature Physics}\ }\textbf {\bibinfo {volume}
  {10}},\ \bibinfo {pages} {218--224} (\bibinfo {year} {2014})}\BibitemShut
  {NoStop}%
\bibitem [{\citenamefont {King}\ \emph {et~al.}(2018)\citenamefont {King},
  \citenamefont {Carrasquilla}, \citenamefont {Raymond}, \citenamefont
  {Ozfidan},\ and\ \citenamefont {Andriyash}}]{king_observation_2018}%
  \BibitemOpen
  \bibfield  {author} {\bibinfo {author} {\bibfnamefont {A.~D.}\ \bibnamefont
  {King}}, \bibinfo {author} {\bibfnamefont {J.}~\bibnamefont {Carrasquilla}},
  \bibinfo {author} {\bibfnamefont {J.}~\bibnamefont {Raymond}}, \bibinfo
  {author} {\bibfnamefont {I.}~\bibnamefont {Ozfidan}}, \ and\ \bibinfo
  {author} {\bibfnamefont {E.~et.~al.}\ \bibnamefont {Andriyash}},\ }\bibfield
  {title} {\enquote {\bibinfo {title} {Observation of topological phenomena in
  a programmable lattice of 1,800 qubits},}\ }\href {\doibase
  10.1038/s41586-018-0410-x} {\bibfield  {journal} {\bibinfo  {journal}
  {Nature}\ }\textbf {\bibinfo {volume} {560}},\ \bibinfo {pages} {456--460}
  (\bibinfo {year} {2018})}\BibitemShut {NoStop}%
\bibitem [{\citenamefont {Venturelli}\ and\ \citenamefont
  {Kondratyev}(2019)}]{venturelli_reverse_2019}%
  \BibitemOpen
  \bibfield  {author} {\bibinfo {author} {\bibfnamefont {D.}~\bibnamefont
  {Venturelli}}\ and\ \bibinfo {author} {\bibfnamefont {A.}~\bibnamefont
  {Kondratyev}},\ }\bibfield  {title} {\enquote {\bibinfo {title} {Reverse
  quantum annealing approach to portfolio optimization problems},}\ }\href
  {\doibase 10.1007/s42484-019-00001-w} {\bibfield  {journal} {\bibinfo
  {journal} {Quantum Machine Intelligence}\ }\textbf {\bibinfo {volume} {1}},\
  \bibinfo {pages} {17--30} (\bibinfo {year} {2019})}\BibitemShut {NoStop}%
\bibitem [{\citenamefont {Venturelli}\ \emph {et~al.}(2018)\citenamefont
  {Venturelli}, \citenamefont {Do}, \citenamefont {Rieffel},\ and\
  \citenamefont {Frank}}]{venturelli_compiling_2018}%
  \BibitemOpen
  \bibfield  {author} {\bibinfo {author} {\bibfnamefont {D.}~\bibnamefont
  {Venturelli}}, \bibinfo {author} {\bibfnamefont {M.}~\bibnamefont {Do}},
  \bibinfo {author} {\bibfnamefont {E.}~\bibnamefont {Rieffel}}, \ and\
  \bibinfo {author} {\bibfnamefont {J.}~\bibnamefont {Frank}},\ }\bibfield
  {title} {\enquote {\bibinfo {title} {Compiling quantum circuits to realistic
  hardware architectures using temporal planners},}\ }\href {\doibase
  10.1088/2058-9565/aaa331} {\bibfield  {journal} {\bibinfo  {journal} {Quantum
  Science and Technology}\ }\textbf {\bibinfo {volume} {3}},\ \bibinfo {pages}
  {025004} (\bibinfo {year} {2018})}\BibitemShut {NoStop}%
\bibitem [{\citenamefont {Neukart}\ \emph {et~al.}(2017)\citenamefont
  {Neukart}, \citenamefont {Compostella}, \citenamefont {Seidel}, \citenamefont
  {von Dollen},\ and\ \citenamefont {Yarkoni}}]{neukart_traffic_2017}%
  \BibitemOpen
  \bibfield  {author} {\bibinfo {author} {\bibfnamefont {F.}~\bibnamefont
  {Neukart}}, \bibinfo {author} {\bibfnamefont {G.}~\bibnamefont
  {Compostella}}, \bibinfo {author} {\bibfnamefont {C.}~\bibnamefont {Seidel}},
  \bibinfo {author} {\bibfnamefont {D.}~\bibnamefont {von Dollen}}, \ and\
  \bibinfo {author} {\bibfnamefont {S.~et.~al.}\ \bibnamefont {Yarkoni}},\
  }\bibfield  {title} {\enquote {\bibinfo {title} {Traffic {Flow}
  {Optimization} {Using} a {Quantum} {Annealer}},}\ }\href {\doibase
  10.3389/fict.2017.00029} {\bibfield  {journal} {\bibinfo  {journal}
  {Frontiers in ICT}\ }\textbf {\bibinfo {volume} {4}} (\bibinfo {year}
  {2017}),\ 10.3389/fict.2017.00029}\BibitemShut {NoStop}%
\bibitem [{\citenamefont {Rosenberg}\ \emph {et~al.}(2016)\citenamefont
  {Rosenberg}, \citenamefont {Haghnegahdar}, \citenamefont {Goddard},
  \citenamefont {Carr},\ and\ \citenamefont {Wu}}]{rosenberg_solving_2016}%
  \BibitemOpen
  \bibfield  {author} {\bibinfo {author} {\bibfnamefont {G.}~\bibnamefont
  {Rosenberg}}, \bibinfo {author} {\bibfnamefont {P.}~\bibnamefont
  {Haghnegahdar}}, \bibinfo {author} {\bibfnamefont {P.}~\bibnamefont
  {Goddard}}, \bibinfo {author} {\bibfnamefont {P.}~\bibnamefont {Carr}}, \
  and\ \bibinfo {author} {\bibfnamefont {K.~et.~al.}\ \bibnamefont {Wu}},\
  }\bibfield  {title} {\enquote {\bibinfo {title} {Solving the {Optimal}
  {Trading} {Trajectory} {Problem} {Using} a {Quantum} {Annealer}},}\ }\href
  {\doibase 10.1109/JSTSP.2016.2574703} {\bibfield  {journal} {\bibinfo
  {journal} {IEEE Journal of Selected Topics in Signal Processing}\ }\textbf
  {\bibinfo {volume} {10}},\ \bibinfo {pages} {1053--1060} (\bibinfo {year}
  {2016})}\BibitemShut {NoStop}%
\bibitem [{\citenamefont {Rieffel}\ \emph {et~al.}(2015)\citenamefont
  {Rieffel}, \citenamefont {Venturelli}, \citenamefont {O'Gorman},
  \citenamefont {Do},\ and\ \citenamefont {Prystay}}]{rieffel_case_2015}%
  \BibitemOpen
  \bibfield  {author} {\bibinfo {author} {\bibfnamefont {E.~G.}\ \bibnamefont
  {Rieffel}}, \bibinfo {author} {\bibfnamefont {D.}~\bibnamefont {Venturelli}},
  \bibinfo {author} {\bibfnamefont {B.}~\bibnamefont {O'Gorman}}, \bibinfo
  {author} {\bibfnamefont {M.~B.}\ \bibnamefont {Do}}, \ and\ \bibinfo {author}
  {\bibfnamefont {E.~M. et.~al.}\ \bibnamefont {Prystay}},\ }\bibfield  {title}
  {\enquote {\bibinfo {title} {A case study in programming a quantum annealer
  for hard operational planning problems},}\ }\href {\doibase
  10.1007/s11128-014-0892-x} {\bibfield  {journal} {\bibinfo  {journal}
  {Quantum Information Processing}\ }\textbf {\bibinfo {volume} {14}},\
  \bibinfo {pages} {1--36} (\bibinfo {year} {2015})}\BibitemShut {NoStop}%
\bibitem [{\citenamefont {Knysh}(2015)}]{knysh_computational_2015}%
  \BibitemOpen
  \bibfield  {author} {\bibinfo {author} {\bibfnamefont {S.}~\bibnamefont
  {Knysh}},\ }\bibfield  {title} {\enquote {\bibinfo {title} {Computational
  {Bottlenecks} of {Quantum} {Annealing}},}\ }\href
  {https://arxiv.org/abs/1506.08608} {\bibfield  {journal} {\bibinfo  {journal}
  {arXiv:1506.08608}\ } (\bibinfo {year} {2015})}\BibitemShut {NoStop}%
\bibitem [{\citenamefont {Young}\ \emph {et~al.}(2010)\citenamefont {Young},
  \citenamefont {Knysh},\ and\ \citenamefont
  {Smelyanskiy}}]{young_first-order_2010}%
  \BibitemOpen
  \bibfield  {author} {\bibinfo {author} {\bibfnamefont {A.~P.}\ \bibnamefont
  {Young}}, \bibinfo {author} {\bibfnamefont {S.}~\bibnamefont {Knysh}}, \ and\
  \bibinfo {author} {\bibfnamefont {V.~N.}\ \bibnamefont {Smelyanskiy}},\
  }\bibfield  {title} {\enquote {\bibinfo {title} {First-{Order} {Phase}
  {Transition} in the {Quantum} {Adiabatic} {Algorithm}},}\ }\href {\doibase
  10.1103/PhysRevLett.104.020502} {\bibfield  {journal} {\bibinfo  {journal}
  {Physiccal Review Letters}\ }\textbf {\bibinfo {volume} {104}},\ \bibinfo
  {pages} {020502} (\bibinfo {year} {2010})}\BibitemShut {NoStop}%
\bibitem [{\citenamefont {Morita}\ and\ \citenamefont
  {Nishimori}(2006)}]{morita_convergence_2006}%
  \BibitemOpen
  \bibfield  {author} {\bibinfo {author} {\bibfnamefont {S.}~\bibnamefont
  {Morita}}\ and\ \bibinfo {author} {\bibfnamefont {H.}~\bibnamefont
  {Nishimori}},\ }\bibfield  {title} {\enquote {\bibinfo {title} {Convergence
  theorems for quantum annealing},}\ }\href {\doibase
  10.1088/0305-4470/39/45/004} {\bibfield  {journal} {\bibinfo  {journal}
  {Journal of Physics A Mathematical General}\ }\textbf {\bibinfo {volume}
  {39}},\ \bibinfo {pages} {13903--13920} (\bibinfo {year} {2006})}\BibitemShut
  {NoStop}%
\bibitem [{\citenamefont {Santoro}\ \emph {et~al.}(2002)\citenamefont
  {Santoro}, \citenamefont {Martonak}, \citenamefont {Tosatti},\ and\
  \citenamefont {Car}}]{santoro_theory_2002}%
  \BibitemOpen
  \bibfield  {author} {\bibinfo {author} {\bibfnamefont {G.~E.}\ \bibnamefont
  {Santoro}}, \bibinfo {author} {\bibfnamefont {R.}~\bibnamefont {Martonak}},
  \bibinfo {author} {\bibfnamefont {E.}~\bibnamefont {Tosatti}}, \ and\
  \bibinfo {author} {\bibfnamefont {R.}~\bibnamefont {Car}},\ }\bibfield
  {title} {\enquote {\bibinfo {title} {Theory of {Quantum} {Annealing} of an
  {Ising} {Spin} {Glass}},}\ }\href {\doibase 10.1126/science.1068774}
  {\bibfield  {journal} {\bibinfo  {journal} {Science}\ }\textbf {\bibinfo
  {volume} {295}},\ \bibinfo {pages} {2427--2430} (\bibinfo {year}
  {2002})}\BibitemShut {NoStop}%
\bibitem [{\citenamefont {Lanting}\ \emph {et~al.}(2017)\citenamefont
  {Lanting}, \citenamefont {King}, \citenamefont {Evert},\ and\ \citenamefont
  {Hoskinson}}]{lanting_experimental_2017}%
  \BibitemOpen
  \bibfield  {author} {\bibinfo {author} {\bibfnamefont {T.}~\bibnamefont
  {Lanting}}, \bibinfo {author} {\bibfnamefont {A.~D.}\ \bibnamefont {King}},
  \bibinfo {author} {\bibfnamefont {B.}~\bibnamefont {Evert}}, \ and\ \bibinfo
  {author} {\bibfnamefont {E.}~\bibnamefont {Hoskinson}},\ }\bibfield  {title}
  {\enquote {\bibinfo {title} {Experimental demonstration of perturbative
  anticrossing mitigation using nonuniform driver {Hamiltonians}},}\ }\href
  {\doibase 10.1103/PhysRevA.96.042322} {\bibfield  {journal} {\bibinfo
  {journal} {Physical Review A}\ }\textbf {\bibinfo {volume} {96}},\ \bibinfo
  {pages} {042322} (\bibinfo {year} {2017})}\BibitemShut {NoStop}%
\bibitem [{\citenamefont {King}\ \emph {et~al.}(2016)\citenamefont {King},
  \citenamefont {Hoskinson}, \citenamefont {Lanting}, \citenamefont
  {Andriyash},\ and\ \citenamefont {Amin}}]{king_degeneracy_2016}%
  \BibitemOpen
  \bibfield  {author} {\bibinfo {author} {\bibfnamefont {A.~D.}\ \bibnamefont
  {King}}, \bibinfo {author} {\bibfnamefont {E.}~\bibnamefont {Hoskinson}},
  \bibinfo {author} {\bibfnamefont {T.}~\bibnamefont {Lanting}}, \bibinfo
  {author} {\bibfnamefont {E.}~\bibnamefont {Andriyash}}, \ and\ \bibinfo
  {author} {\bibfnamefont {M.~H.}\ \bibnamefont {Amin}},\ }\bibfield  {title}
  {\enquote {\bibinfo {title} {Degeneracy, degree, and heavy tails in quantum
  annealing},}\ }\href {\doibase 10.1103/PhysRevA.93.052320} {\bibfield
  {journal} {\bibinfo  {journal} {Physical Review A}\ }\textbf {\bibinfo
  {volume} {93}},\ \bibinfo {pages} {052320} (\bibinfo {year}
  {2016})}\BibitemShut {NoStop}%
\bibitem [{\citenamefont {Steiger}\ \emph {et~al.}(2015)\citenamefont
  {Steiger}, \citenamefont {Ronnow},\ and\ \citenamefont
  {Troyer}}]{steiger_heavy_2015}%
  \BibitemOpen
  \bibfield  {author} {\bibinfo {author} {\bibfnamefont {D.~S.}\ \bibnamefont
  {Steiger}}, \bibinfo {author} {\bibfnamefont {T.~F.}\ \bibnamefont {Ronnow}},
  \ and\ \bibinfo {author} {\bibfnamefont {M.}~\bibnamefont {Troyer}},\
  }\bibfield  {title} {\enquote {\bibinfo {title} {Heavy {Tails} in the
  {Distribution} of {Time} to {Solution} for {Classical} and {Quantum}
  {Annealing}},}\ }\href {\doibase 10.1103/PhysRevLett.115.230501} {\bibfield
  {journal} {\bibinfo  {journal} {Physical Review Letters}\ }\textbf {\bibinfo
  {volume} {115}},\ \bibinfo {pages} {230501} (\bibinfo {year}
  {2015})}\BibitemShut {NoStop}%
\bibitem [{\citenamefont {Dickson}\ \emph {et~al.}(2013)\citenamefont
  {Dickson}, \citenamefont {Johnson}, \citenamefont {Amin}, \citenamefont
  {Harris},\ and\ \citenamefont {Altomare}}]{dickson_thermally_2013}%
  \BibitemOpen
  \bibfield  {author} {\bibinfo {author} {\bibfnamefont {N.~G.}\ \bibnamefont
  {Dickson}}, \bibinfo {author} {\bibfnamefont {M.~W.}\ \bibnamefont
  {Johnson}}, \bibinfo {author} {\bibfnamefont {M.~H.}\ \bibnamefont {Amin}},
  \bibinfo {author} {\bibfnamefont {R.}~\bibnamefont {Harris}}, \ and\ \bibinfo
  {author} {\bibfnamefont {F.~et.~al.}\ \bibnamefont {Altomare}},\ }\bibfield
  {title} {\enquote {\bibinfo {title} {Thermally assisted quantum annealing of
  a 16-qubit problem},}\ }\href {\doibase 10.1038/ncomms2920} {\bibfield
  {journal} {\bibinfo  {journal} {Nature Communications}\ }\textbf {\bibinfo
  {volume} {4}},\ \bibinfo {pages} {1903} (\bibinfo {year} {2013})}\BibitemShut
  {NoStop}%
\bibitem [{\citenamefont {Altshuler}\ \emph {et~al.}(2010)\citenamefont
  {Altshuler}, \citenamefont {Krovi},\ and\ \citenamefont
  {Roland}}]{altshuler_anderson_2010}%
  \BibitemOpen
  \bibfield  {author} {\bibinfo {author} {\bibfnamefont {B.}~\bibnamefont
  {Altshuler}}, \bibinfo {author} {\bibfnamefont {H.}~\bibnamefont {Krovi}}, \
  and\ \bibinfo {author} {\bibfnamefont {J.}~\bibnamefont {Roland}},\
  }\bibfield  {title} {\enquote {\bibinfo {title} {Anderson localization makes
  adiabatic quantum optimization fail},}\ }\href {\doibase
  10.1073/pnas.1002116107} {\bibfield  {journal} {\bibinfo  {journal}
  {Proceedings of the National Academy of Sciences}\ }\textbf {\bibinfo
  {volume} {107}},\ \bibinfo {pages} {12446--12450} (\bibinfo {year}
  {2010})}\BibitemShut {NoStop}%
\bibitem [{\citenamefont {Keck}\ \emph {et~al.}(2017)\citenamefont {Keck},
  \citenamefont {Montangero}, \citenamefont {Santoro}, \citenamefont {Fazio},\
  and\ \citenamefont {Rossini}}]{keck_dissipation_2017}%
  \BibitemOpen
  \bibfield  {author} {\bibinfo {author} {\bibfnamefont {M.}~\bibnamefont
  {Keck}}, \bibinfo {author} {\bibfnamefont {S.}~\bibnamefont {Montangero}},
  \bibinfo {author} {\bibfnamefont {G.~E.}\ \bibnamefont {Santoro}}, \bibinfo
  {author} {\bibfnamefont {R.}~\bibnamefont {Fazio}}, \ and\ \bibinfo {author}
  {\bibfnamefont {D.}~\bibnamefont {Rossini}},\ }\bibfield  {title} {\enquote
  {\bibinfo {title} {Dissipation in adiabatic quantum computers: lessons from
  an exactly solvable model},}\ }\href {\doibase 10.1088/1367-2630/aa8cef}
  {\bibfield  {journal} {\bibinfo  {journal} {New Journal of Physics}\ }\textbf
  {\bibinfo {volume} {19}},\ \bibinfo {pages} {113029} (\bibinfo {year}
  {2017})}\BibitemShut {NoStop}%
\bibitem [{\citenamefont {Deng}\ \emph {et~al.}(2013)\citenamefont {Deng},
  \citenamefont {Averin}, \citenamefont {Amin},\ and\ \citenamefont
  {Smith}}]{deng_decoherence_2013}%
  \BibitemOpen
  \bibfield  {author} {\bibinfo {author} {\bibfnamefont {Q.}~\bibnamefont
  {Deng}}, \bibinfo {author} {\bibfnamefont {D.~V.}\ \bibnamefont {Averin}},
  \bibinfo {author} {\bibfnamefont {M.~H.}\ \bibnamefont {Amin}}, \ and\
  \bibinfo {author} {\bibfnamefont {P.}~\bibnamefont {Smith}},\ }\bibfield
  {title} {\enquote {\bibinfo {title} {Decoherence induced deformation of the
  ground state in adiabatic quantum computation},}\ }\href {\doibase
  10.1038/srep01479} {\bibfield  {journal} {\bibinfo  {journal} {Scientific
  Reports}\ }\textbf {\bibinfo {volume} {3}},\ \bibinfo {pages} {1479}
  (\bibinfo {year} {2013})}\BibitemShut {NoStop}%
\bibitem [{\citenamefont {Boixo}\ \emph {et~al.}(2016)\citenamefont {Boixo},
  \citenamefont {Smelyanskiy}, \citenamefont {Shabani}, \citenamefont {Isakov},
  \citenamefont {Dykman},\ and\ \citenamefont
  {{others}}}]{boixo_computational_2016}%
  \BibitemOpen
  \bibfield  {author} {\bibinfo {author} {\bibfnamefont {S.}~\bibnamefont
  {Boixo}}, \bibinfo {author} {\bibfnamefont {V.~N.}\ \bibnamefont
  {Smelyanskiy}}, \bibinfo {author} {\bibfnamefont {A.}~\bibnamefont
  {Shabani}}, \bibinfo {author} {\bibfnamefont {S.~V.}\ \bibnamefont {Isakov}},
  \bibinfo {author} {\bibfnamefont {M.}~\bibnamefont {Dykman}}, \ and\ \bibinfo
  {author} {\bibnamefont {{others}}},\ }\bibfield  {title} {\enquote {\bibinfo
  {title} {Computational multiqubit tunnelling in programmable quantum
  annealers},}\ }\href {\doibase 10.1038/ncomms10327} {\bibfield  {journal}
  {\bibinfo  {journal} {Nature communications}\ }\textbf {\bibinfo {volume}
  {7}},\ \bibinfo {pages} {10327} (\bibinfo {year} {2016})}\BibitemShut
  {NoStop}%
\bibitem [{\citenamefont {Amin}\ \emph {et~al.}(2009)\citenamefont {Amin},
  \citenamefont {Truncik},\ and\ \citenamefont {Averin}}]{amin_role_2009}%
  \BibitemOpen
  \bibfield  {author} {\bibinfo {author} {\bibfnamefont {M.~H.~S.}\
  \bibnamefont {Amin}}, \bibinfo {author} {\bibfnamefont {C.~J.~S.}\
  \bibnamefont {Truncik}}, \ and\ \bibinfo {author} {\bibfnamefont {D.~V.}\
  \bibnamefont {Averin}},\ }\bibfield  {title} {\enquote {\bibinfo {title}
  {Role of single-qubit decoherence time in adiabatic quantum computation},}\
  }\href {\doibase 10.1103/PhysRevA.80.022303} {\bibfield  {journal} {\bibinfo
  {journal} {Physical Review A}\ }\textbf {\bibinfo {volume} {80}},\ \bibinfo
  {pages} {022303} (\bibinfo {year} {2009})}\BibitemShut {NoStop}%
\bibitem [{\citenamefont {Lieb}\ \emph {et~al.}(1961)\citenamefont {Lieb},
  \citenamefont {Schultz},\ and\ \citenamefont {Mattis}}]{lieb_two_1961}%
  \BibitemOpen
  \bibfield  {author} {\bibinfo {author} {\bibfnamefont {E.}~\bibnamefont
  {Lieb}}, \bibinfo {author} {\bibfnamefont {T.}~\bibnamefont {Schultz}}, \
  and\ \bibinfo {author} {\bibfnamefont {D.}~\bibnamefont {Mattis}},\
  }\bibfield  {title} {\enquote {\bibinfo {title} {Two soluble models of an
  antiferromagnetic chain},}\ }\href {\doibase 10.1016/0003-4916(61)90115-4}
  {\bibfield  {journal} {\bibinfo  {journal} {Annals of Physics}\ }\textbf
  {\bibinfo {volume} {16}},\ \bibinfo {pages} {407--466} (\bibinfo {year}
  {1961})}\BibitemShut {NoStop}%
\bibitem [{\citenamefont {Dziarmaga}(2005)}]{dziarmaga_dynamics_2005}%
  \BibitemOpen
  \bibfield  {author} {\bibinfo {author} {\bibfnamefont {J.}~\bibnamefont
  {Dziarmaga}},\ }\bibfield  {title} {\enquote {\bibinfo {title} {Dynamics of a
  {Quantum} {Phase} {Transition}: {Exact} {Solution} of the {Quantum} {Ising}
  {Model}},}\ }\href {\doibase 10.1103/PhysRevLett.95.245701} {\bibfield
  {journal} {\bibinfo  {journal} {Physical Review Letters}\ }\textbf {\bibinfo
  {volume} {95}},\ \bibinfo {pages} {245701} (\bibinfo {year}
  {2005})}\BibitemShut {NoStop}%
\bibitem [{\citenamefont {Bravyi}\ \emph {et~al.}(2010)\citenamefont {Bravyi},
  \citenamefont {Hastings},\ and\ \citenamefont
  {Michalakis}}]{bravyi_topological_2010}%
  \BibitemOpen
  \bibfield  {author} {\bibinfo {author} {\bibfnamefont {Sergey}\ \bibnamefont
  {Bravyi}}, \bibinfo {author} {\bibfnamefont {Matthew~B.}\ \bibnamefont
  {Hastings}}, \ and\ \bibinfo {author} {\bibfnamefont {Spyridon}\ \bibnamefont
  {Michalakis}},\ }\bibfield  {title} {\enquote {\bibinfo {title} {Topological
  quantum order: Stability under local perturbations},}\ }\href {\doibase
  10.1063/1.3490195} {\bibfield  {journal} {\bibinfo  {journal} {Journal of
  Mathematical Physics}\ }\textbf {\bibinfo {volume} {51}},\ \bibinfo {pages}
  {093512} (\bibinfo {year} {2010})},\ \Eprint
  {http://arxiv.org/abs/https://doi.org/10.1063/1.3490195}
  {https://doi.org/10.1063/1.3490195} \BibitemShut {NoStop}%
\bibitem [{\citenamefont {Chen}\ \emph {et~al.}(2010)\citenamefont {Chen},
  \citenamefont {Gu},\ and\ \citenamefont {Wen}}]{chen_local_2010}%
  \BibitemOpen
  \bibfield  {author} {\bibinfo {author} {\bibfnamefont {X.}~\bibnamefont
  {Chen}}, \bibinfo {author} {\bibfnamefont {Z.~C.}\ \bibnamefont {Gu}}, \ and\
  \bibinfo {author} {\bibfnamefont {X.~G.}\ \bibnamefont {Wen}},\ }\bibfield
  {title} {\enquote {\bibinfo {title} {Local unitary transformation, long-range
  quantum entanglement, wave function renormalization, and topological
  order},}\ }\href {\doibase 10.1103/PhysRevB.82.155138} {\bibfield  {journal}
  {\bibinfo  {journal} {Physical Review B}\ }\textbf {\bibinfo {volume} {82}},\
  \bibinfo {pages} {155138} (\bibinfo {year} {2010})}\BibitemShut {NoStop}%
\bibitem [{\citenamefont {Iorgov}(2011)}]{iorgov_form_2011}%
  \BibitemOpen
  \bibfield  {author} {\bibinfo {author} {\bibfnamefont {Nikolai}\ \bibnamefont
  {Iorgov}},\ }\bibfield  {title} {\enquote {\bibinfo {title} {Form factors of
  the finite {quantumXY}-chain},}\ }\href {\doibase
  10.1088/1751-8113/44/33/335005} {\bibfield  {journal} {\bibinfo  {journal}
  {Journal of Physics A: Mathematical and Theoretical}\ }\textbf {\bibinfo
  {volume} {44}},\ \bibinfo {pages} {335005} (\bibinfo {year}
  {2011})}\BibitemShut {NoStop}%
\bibitem [{\citenamefont {Marshall}\ \emph {et~al.}(2017)\citenamefont
  {Marshall}, \citenamefont {Rieffel},\ and\ \citenamefont
  {Hen}}]{marshall_thermalization_2017}%
  \BibitemOpen
  \bibfield  {author} {\bibinfo {author} {\bibfnamefont {J.}~\bibnamefont
  {Marshall}}, \bibinfo {author} {\bibfnamefont {E.~G.}\ \bibnamefont
  {Rieffel}}, \ and\ \bibinfo {author} {\bibfnamefont {I.}~\bibnamefont
  {Hen}},\ }\bibfield  {title} {\enquote {\bibinfo {title} {Thermalization,
  {Freeze}-out, and {Noise}: {Deciphering} {Experimental} {Quantum}
  {Annealers}},}\ }\href {\doibase 10.1103/PhysRevApplied.8.064025} {\bibfield
  {journal} {\bibinfo  {journal} {Physical Review Applied}\ }\textbf {\bibinfo
  {volume} {8}},\ \bibinfo {pages} {064025} (\bibinfo {year}
  {2017})}\BibitemShut {NoStop}%
\bibitem [{\citenamefont {Amin}(2015)}]{amin_searching_2015}%
  \BibitemOpen
  \bibfield  {author} {\bibinfo {author} {\bibfnamefont {M.~H.}\ \bibnamefont
  {Amin}},\ }\bibfield  {title} {\enquote {\bibinfo {title} {Searching for
  quantum speedup in quasistatic quantum annealers},}\ }\href {\doibase
  10.1103/PhysRevA.92.052323} {\bibfield  {journal} {\bibinfo  {journal}
  {Physical Review A}\ }\textbf {\bibinfo {volume} {92}},\ \bibinfo {pages}
  {052323} (\bibinfo {year} {2015})}\BibitemShut {NoStop}%
\bibitem [{\citenamefont {Johnson}\ \emph {et~al.}(2011)\citenamefont
  {Johnson}, \citenamefont {Amin}, \citenamefont {Gildert}, \citenamefont
  {Lanting},\ and\ \citenamefont {Hamze}}]{johnson_quantum_2011}%
  \BibitemOpen
  \bibfield  {author} {\bibinfo {author} {\bibfnamefont {M.~W.}\ \bibnamefont
  {Johnson}}, \bibinfo {author} {\bibfnamefont {M.~H.~S.}\ \bibnamefont
  {Amin}}, \bibinfo {author} {\bibfnamefont {S.}~\bibnamefont {Gildert}},
  \bibinfo {author} {\bibfnamefont {T.}~\bibnamefont {Lanting}}, \ and\
  \bibinfo {author} {\bibfnamefont {F.~et.~al.}\ \bibnamefont {Hamze}},\
  }\bibfield  {title} {\enquote {\bibinfo {title} {Quantum annealing with
  manufactured spins},}\ }\href {\doibase 10.1038/nature10012} {\bibfield
  {journal} {\bibinfo  {journal} {Nature}\ }\textbf {\bibinfo {volume} {473}},\
  \bibinfo {pages} {194--198} (\bibinfo {year} {2011})}\BibitemShut {NoStop}%
\bibitem [{\citenamefont {Palmer}(2007)}]{palmer_thermodynamic_2007}%
  \BibitemOpen
  \bibfield  {author} {\bibinfo {author} {\bibfnamefont {J.}~\bibnamefont
  {Palmer}},\ }\bibfield  {title} {\enquote {\bibinfo {title} {The
  {Thermodynamic} {Limit}},}\ }in\ \href {\doibase 10.1007/978-0-8176-4620-2_1}
  {\emph {\bibinfo {booktitle} {Planar {Ising} {Correlations}}}},\ \bibinfo
  {series and number} {Progress in {Mathematical} {Physics}},\ \bibinfo
  {editor} {edited by\ \bibinfo {editor} {\bibfnamefont {J.}~\bibnamefont
  {Palmer}}}\ (\bibinfo  {publisher} {Birkhauser Boston},\ \bibinfo {address}
  {Boston, MA},\ \bibinfo {year} {2007})\ pp.\ \bibinfo {pages}
  {1--61}\BibitemShut {NoStop}%
\bibitem [{\citenamefont {Sachdev}(2000)}]{sachdev_quantum_2000}%
  \BibitemOpen
  \bibfield  {author} {\bibinfo {author} {\bibfnamefont {S.}~\bibnamefont
  {Sachdev}},\ }\href {\doibase 10.1017/CBO9780511622540} {\enquote {\bibinfo
  {title} {Quantum {Phase} {Transitions}},}\ } (\bibinfo {year}
  {2000})\BibitemShut {NoStop}%
\bibitem [{\citenamefont {Costello}(2011)}]{costello_renormalization_2011}%
  \BibitemOpen
  \bibfield  {author} {\bibinfo {author} {\bibfnamefont {K.}~\bibnamefont
  {Costello}},\ }\href@noop {} {\emph {\bibinfo {title} {Renormalization and
  effective field theory}}},\ \bibinfo {number} {170}\ (\bibinfo  {publisher}
  {American Mathematical Soc.},\ \bibinfo {year} {2011})\BibitemShut {NoStop}%
\bibitem [{\citenamefont {Schultz}\ \emph {et~al.}(1964)\citenamefont
  {Schultz}, \citenamefont {Mattis},\ and\ \citenamefont
  {Lieb}}]{schultz_two-dimensional_1964}%
  \BibitemOpen
  \bibfield  {author} {\bibinfo {author} {\bibfnamefont {T.}~\bibnamefont
  {Schultz}}, \bibinfo {author} {\bibfnamefont {D.}~\bibnamefont {Mattis}}, \
  and\ \bibinfo {author} {\bibfnamefont {E.}~\bibnamefont {Lieb}},\ }\bibfield
  {title} {\enquote {\bibinfo {title} {Two-{Dimensional} {Ising} {Model} as a
  {Soluble} {Problem} of {Many} {Fermions}},}\ }\href {\doibase
  10.1103/RevModPhys.36.856} {\bibfield  {journal} {\bibinfo  {journal}
  {Reviews of Modern Physics}\ }\textbf {\bibinfo {volume} {36}},\ \bibinfo
  {pages} {856--871} (\bibinfo {year} {1964})}\BibitemShut {NoStop}%
\bibitem [{\citenamefont {Yang}(1952)}]{yang_spontaneous_1952}%
  \BibitemOpen
  \bibfield  {author} {\bibinfo {author} {\bibfnamefont {C.~N.}\ \bibnamefont
  {Yang}},\ }\bibfield  {title} {\enquote {\bibinfo {title} {The {Spontaneous}
  {Magnetization} of a {Two}-{Dimensional} {Ising} {Model}},}\ }\href {\doibase
  10.1103/PhysRev.85.808} {\bibfield  {journal} {\bibinfo  {journal} {Physical
  Review}\ }\textbf {\bibinfo {volume} {85}},\ \bibinfo {pages} {808--816}
  (\bibinfo {year} {1952})}\BibitemShut {NoStop}%
\end{thebibliography}

%

\appendix

\section{Exact diagonalization of the Frustrated Ring} \label{app:low}
In this Appendix we systematically and rigorously solve for the bound states of the Frustrated Ring,
deriving their crossing properties and quantum numbers. Our starting point is the free-fermion representation of the one-dimensional transverse-field Ising spin glass:
\begin{equation}
\hat{H}_0^\pm(t)=-i \sum_{j=1}^\mN J_j^\pm\, \hat{\gamma}_{2j}\hat{\gamma}_{2j+1}
+i B(t)\sum_{j=1}^\mN\hat{\gamma}_{2j-1}\hat{\gamma}_{2j}\ ,\label{eq:fmodel_pair_of_models}
\end{equation}
where $J^\pm_j\equiv J_j$ for $j\neq \mN$ and  $J_\mN^\pm \equiv \pm
J_\mN$. $\hat{H}^\pm_0(t)$ denotes the quantum annealing Hamiltonian restricted to the sector with an even and odd number of fermions respectively. Note that the representation \eqref{eq:fmodel_pair_of_models} of the Frustrated Ring as a pair of free fermion models
is a two-fold redundant description, with each free fermion theory being
valid only in its corresponding parity sector (the sector with an even number of fermions is called the {\it Ramond sector}, and the sector with an odd number is called the {\it Neveu-Schwarz sector}).

The vector space of all quadratic polynomials in Majorana fermions, i.e. all expressions of the form
\begin{align}
    \hat{H}=\sum_{1\leq i<j\leq 2\mN}\lambda_{ij}\hat{\gamma}_i\hat{\gamma}_j,
\end{align}
form a closed Lie algebra isomorphic to $ \mathfrak{spin}_{2\mN}$, i.e. the Lie algebra of the Spin group $\text{Spin}(2\mN)$. It is well-known that this Lie algebra is isomorphic to $\mathfrak{so}_{2\mN}$, i.e. the Lie algebra of real, $2\mN\times 2\mN$ antisymmetric matrices. The isomorphism is given by
\begin{align}
    \hat{H}=\sum_{1\leq i<j\leq 2\mN}\lambda_{ij}\hat{\gamma}_i\hat{\gamma}_j~\mapsto~\mathbb{H}_\text{BdG}\equiv  \begin{pmatrix}&\lambda_{ij}\\-\lambda_{ij}&\end{pmatrix}.
\end{align}
This maps a free-fermion model to its corresponding {\it Bogoliubov de-Gennes} (BdG) Hamiltonian. The Bogoliubov de-Gennes Hamiltonian is a much more lucid representation of a free-fermion model, because the physical data of interest, namely, the quasiparticle operators expressed in the Majorana basis, are given by the eigenvectors of this matrix, and the quasiparticle dispersion corresponds to the eigenvalues. The BdG representation of the one-dimensional transverse-field Ising spin glass is
\begin{align}
\mathbb{H}_\text{BdG}^\pm& = \begin{pmatrix}
 0&B(t)&&&&\pm J_{\mN}\\ -B(t)&0&J_1&&&\\ &-J_1&&\ddots&\\ &&\ddots&&J_{\mN-1}&\\ &&&-J_{\mN-1}&0&B(t)\\\mp J_{\mN}&&&&-B(t)&0\end{pmatrix}\label{eq:quadratic_ham}
\end{align}
In the particular case of the Frustrated Ring, after a change-of-basis
\begin{equation}
\hat{\Gamma}_{2j}\equiv i\hat{\gamma}_{2j},\quad \hat{\Gamma}_{2j-1}\equiv \hat{\gamma}_{2j-1} \ ,\label{eq:fmodel_BdG}
\end{equation}
the matrices specified by \eqref{eq:quadratic_ham}  admit a $\mathbb{Z}_2$ symmetry. It reverses the order of
the basis elements, corresponding to the reflection symmetry of the Frustrated Ring
($J_j\to J_{\mN-(j-1)}$). As is standard, the symmetry splits our eigenvalue problem into two subspaces, indexed by the eigenvalue under reversion (which we denote by $\mu$). The {\it symmetric subspace} (i.e. $\mu=1$) is $\mN$-dimensional, with basis 
\begin{align}
    \hat{\Gamma}_j^+ &\equiv \hat{\Gamma}_j + \hat{\Gamma}_{2\mN -(j-1)} \ ,\label{eq:gammamu1}
\end{align}
whereas the {\it antisymmetric subspace} (i.e. $\mu = -1$) is also $\mN$-dimensional, with basis
\begin{align}
    \hat{\Gamma}_j^- &\equiv \hat{\Gamma}_j - \hat{\Gamma}_{2\mN -(j-1)}.\label{eq:gammamu2}
\end{align}
Under this splitting, the pair of
BdG Hamiltonians \eqref{eq:quadratic_ham} take the block-diagonal form
\begin{equation}
\mathbb{H}_\text{BdG}^\sigma \sim \begin{pmatrix}\mathbb{H}_\text{BdG}^{\sigma,+}&\mathbb{0}_{\mN\times \mN}\\ \mathbb{0}_{\mN\times \mN}& \mathbb{H}_\text{BdG}^{\sigma,-}\end{pmatrix}\ ,
\end{equation}
where
\begin{equation}
\mathbb{H}_\text{BdG}^{\sigma,\mu} = \begin{pmatrix}-\sigma \mu
J_R&B(t)&&&\\B(t)&0&-J&&\\ &-J&&\ddots&\\ &&\ddots&0&-J_L\\ &&&-J_L&\mu
B(t)\end{pmatrix} \ .\label{eq:four_tridiagonals}
\end{equation}
Where, here, $\sigma=\pm 1$ denotes fermion parity. As outlined in the main body of the text, we now find exact low-energy
solutions to the BdG equations in the limit $\mN \to \infty$. In this limit,
we can treat the BdG boundstate problem as a pair of semi-infinite bound state problems,
\begin{align}
    \mathbb{H}^{\sigma,\mu}_\text{BdG}\psi_{\sigma,\mu}= \epsilon_{\sigma,\mu} \psi_{\sigma,\mu}.\label{eq:eigenvalue_problem}
\end{align}
The first such problem looks for a boundstate localized at the right end of the graph:
\begin{align}
\psi^R_{\sigma\mu}&\equiv \begin{pmatrix}\alpha_{\sigma \mu}\\ \beta_{\sigma \mu} \\e^{-\kappa_{\sigma\mu}}\alpha_{\sigma \mu}\\e^{-\kappa_{\sigma\mu}}\beta_{\sigma \mu}\\e^{-2\kappa_{\sigma\mu}}\alpha_{\sigma \mu}\\e^{-2\kappa_{\sigma\mu}}\beta_{\sigma \mu}\\\vdots \end{pmatrix} \ ,\label{eq:evancescent_R}
\end{align}
where here, $\alpha_{\sigma\mu},\beta_{\sigma\mu}$ are written so that it is clear that they only depend on $\sigma$ and $\mu$ through their product. This is because, in the limit $\mN\to\infty$, the decay of the mode \eqref{eq:evancescent_R}  means that the matrix element $\mu B(t)$ in the bottom-right corner of \eqref{eq:four_tridiagonals} does not enter the eigenvalue problem \eqref{eq:eigenvalue_problem}. Similarly, we can solve the left-localized eigenvalue problem:
\begin{align}
\psi^L_{\mu}&\equiv \begin{pmatrix} \alpha_{\mu}\\ \beta_{\mu}\\e^{+\kappa_{\mu}}\alpha_{ \mu}\\e^{+\kappa_{\mu}}\beta_{\mu} \\e^{+2\kappa_{\mu}}\alpha_{ \mu}\\e^{+2\kappa_{\mu}}\beta_{ \mu}\\ \vdots \end{pmatrix}.\label{eq:evancescent_L}
\end{align}
Again, here, $\alpha_{\mu},\beta_{\mu}$ are written so that it is clear that they are independent of $\sigma$. This is because, in the limit $\mN\to\infty$, the decay of the mode \eqref{eq:evancescent_L}  means that the only matrix element containing $\sigma$ (the matrix element in the upper-left corner of \eqref{eq:four_tridiagonals}) does not enter the eigenvalue problem \eqref{eq:eigenvalue_problem}.

Substituting the ans\"{a}tze (\ref{eq:evancescent_R}-\ref{eq:evancescent_L}) into the eigenvalue problem \eqref{eq:eigenvalue_problem}, and letting $$e^{-\lambda}\equiv \alpha/\beta$$ parametrize the ratio of $\alpha$ to $\beta$, we get two main types of equations: bulk conditions, and boundary conditions. The equations in the bulk give
\begin{align}
    \epsilon&=e^{-\lambda}(B-Je^{+\kappa}),\label{eq:bulk_eq1} \ ,\\
    \epsilon&=e^{+\lambda}(B-Je^{-\kappa}) \ .\label{eq:bulk_eq2}
\end{align}
From Eqs. (\ref{eq:bulk_eq1}-\ref{eq:bulk_eq2}), we get the dispersion relation, as well as (after some hyperbolic trigonometry) some useful bulk identities involving $\lambda$:
\begin{align}
    \epsilon^2&=J^2+B^2-2JB\cosh\kappa  \ ,\label{eq:dispersion}\\
    B^2&=\epsilon^2+J^2-2J\epsilon \cosh(\lambda-\kappa) \ ,\label{eq:penult3}\\
    &\tanh \lambda = \frac{J\sinh \kappa}{J\cosh\kappa - B} \ .\label{eq:lambda}
\end{align}
For the boundary conditions, we get different conditions at opposite ends (as would be expected; c.f. \eqref{eq:four_tridiagonals}): for the right-localized boundstate ans\"{a}tz \eqref{eq:evancescent_R}, the boundary conditions are
\begin{equation}
Be^{\lambda} = \epsilon+ \sigma\mu J_R \ ,\label{eq:bcs_R}
\end{equation}
whereas, for the left-localized boundstate ans\"{a}tz \eqref{eq:evancescent_L}, the boundary conditions are
\begin{equation}
Be^{-\lambda} = \epsilon-\frac{ J_L^2}{\epsilon-\mu B} 
\ .\label{eq:bcs_L}
\end{equation}
The right-boundary condition \eqref{eq:bcs_R} generically
yields one solution, $\psi^R_{\sigma\mu}$ for each value of the product $\sigma\mu$. When $\sigma\mu =+1$, we denote this solution as $\psi^R_+$, and when $\sigma\mu = -1$, we denote this solution as $\psi^R_-$. For $B\to 0$, these solutions have the limiting form
\begin{align}
    \psi^R_{\pm} \underset{B\to 0}{\sim} \begin{pmatrix}1\\0\\0\\\vdots\end{pmatrix} \ .\label{eq:evanescent_mode_limiting_form_R}
\end{align}
Similarly, the left-boundary condition \eqref{eq:bcs_L} generically
yields two solutions for each value of $\mu$. For $B\to 0$, the solution whose eigenvalue is least in magnitude limits to
\begin{align}
    \psi^{L}_\mu\underset{B\to 0}{\sim}\begin{pmatrix}\vdots\\ 0\\ 0\\ 1\\  \mu
    \end{pmatrix} \ ,\label{eq:evanescent_mode_limiting_form_L1}
\end{align}
whereas, for $B\to 0$, the solution whose eigenvalue is greatest in magnitude limits to
\begin{align}
\psi^{L'}_\mu\underset{B\to 0}{\sim}\begin{pmatrix}\vdots\\ 0\\ 0\\ 1\\ -\mu
    \end{pmatrix}\ .\label{eq:evanescent_mode_limiting_form_L2}
\end{align}
In summary, putting together both the left- and right-eigenvalue problems, there are {\it six} subgap states in total for our spin glass benchmark.

To write down the quasiparticle/quasihole excitations corresponding to a given boundstate, we simply contract the boundstate vector with the Gamma matrices. This is succinctly captured by the Feynman slash notation, which is the standard physics notation for such expressions:
\begin{align}
    \not{\psi}\equiv \sum_{j=1}^{2\mN} \psi_j \hat{\gamma}_j \ .
\end{align}
with $\psi_j$ the components of the BdG eigenvector in the original basis $\hat{\gamma}_j$ (c.f. \eqref{eq:quadratic_ham}). Having enumerated the subgap states in equations (\ref{eq:evanescent_mode_limiting_form_R}-\ref{eq:evanescent_mode_limiting_form_L2}), we must now physically identify them as localized excitations in our spin glass. To do this, we define
\begin{align}
    \hat{H}^\pm_0\underset{B\to 0}{\sim}\sum_{j=1}^\mN J^\pm_j \hat{c}_{j,\pm}^\dag\hat{c}_{j,\pm}.
\end{align}
Cross-matching with Eqs. (\ref{eq:evanescent_mode_limiting_form_R}-\ref{eq:evanescent_mode_limiting_form_L2}), we get, in the $\mu = +1$ sector:
\begin{align}
    \not{\psi}_\pm^R \underset{B\to 0}{\propto} \hat{c}_{2n+1,\pm}^\dag,\label{eq:R_BdG_mu_positive}\\
    \not{\psi}_+^L \underset{B\to 0}{\propto} \hat{c}_{n,\pm}+\hat{c}_{n+1,\pm},\label{eq:L_BdG1_mu_positive}\\
    \not{\psi}_+^{L'} \underset{B\to 0}{\propto} \hat{c}_{n,\pm}^\dag-\hat{c}_{n+1,\pm}^\dag,\label{eq:L_BdG2_mu_positive}
\end{align}
where here, we have factored $\mN\equiv 2n+1$. Similarly, in the $\mu= -1$ sector, we get
\begin{align}
    \not{\psi}_\pm^R \underset{B\to 0}{\propto} \hat{c}_{2n+1,\mp},\label{eq:R_BdG_mu_negative}\\
    \not{\psi}_-^L \underset{B\to 0}{\propto} \hat{c}_{n,\pm}^\dag+\hat{c}_{n+1,\pm}^\dag,\label{eq:L_BdG1_mu_negative}\\
    \not{\psi}_-^{L'} \underset{B\to 0}{\propto} \hat{c}_{n,\pm}-\hat{c}_{n+1,\pm}.\label{eq:L_BdG2_mu_negative}
\end{align}
This concludes the main exposition of the boundstate problem for the Frustrated Ring. In the following sections, we will apply this knowledge to derive all identities used in the main text.

\subsubsection{Definitions of  spin-glass excitations used in the main text}
Our analysis thus culminates in our first result, which is to give a precise definition of the excitations used in the main text: the identities (\ref{eq:R_BdG_mu_positive}-\ref{eq:L_BdG2_mu_negative}) derived in the previous subsection physically motivate the following definitions at $B=0$:
\begin{align}
    c_{R,\pm}^\dag|_{B=0} &\equiv \hat{c}_{2n+1,\pm}^\dag \ , \label{eq:naive_defs_zeroB1}\\
    c_{L,\pm}^\dag|_{B=0} &\equiv \hat{c}_{n,\pm}^\dag +\hat{c}_{n+1,\pm}^\dag \ ,\\
    c_{L',\pm}^\dag|_{B=0} &\equiv \hat{c}_{n,\pm}^\dag -\hat{c}_{n+1,\pm}^\dag \ . \label{eq:naive_defs_zeroB3}
\end{align}
We can now use (\ref{eq:R_BdG_mu_positive}-\ref{eq:L_BdG2_mu_negative}) to analytically continue the above definitions to non-zero transverse-field $B \neq 0$, by simply tracking the BdG boundstates as they evolve:
\begin{align}
    \hat{c}_{R,\pm}^\dag(B)&:=\sum_j (\psi^R_+)_j\hat{\Gamma}_j^\pm\label{eq:text1} \ ,\\
    \hat{c}_{L,\pm}^\dag(B)&:=\sum_j (\psi^L_-)_j\hat{\Gamma}_j^-\label{eq:text2} \ ,\\
    \hat{c}_{L',\pm}^\dag(B)&:=\sum_j (\psi^{L'}_+)_j\hat{\Gamma}_j^+ \ .\label{eq:text3}
\end{align}
Eqs. (\ref{eq:text1}-\ref{eq:text3}), along with the definition of the BdG eigenvectors (\ref{eq:evancescent_R}-\ref{eq:evancescent_L}), give a mathematically precise notion of these excitations existing throughout the spin glass phase $B<B_c$.\\

Armed with Eqs. (\ref{eq:text1}-\ref{eq:text3}), we can now prove that, in our spin glass benchmark, the crossing in the odd-parity sector is exact, whereas the even-parity crossing is not. It suffices to look at the positive-energy crossing, as states with energies of opposite sign cannot cross in the spin glass phase $B<B_c$ (by definition). For $\sigma = +1$, the only right-localized boundstate with positive eigenvalue is the quasihole
\begin{align}
    \hat{c}_{R,+}.
\end{align}
By taking the adjoint of \eqref{eq:text1}, we see that this excitation consists only of $\mu=-1$ Gamma matrices, as (c.f. \ref{eq:fmodel_BdG}-\ref{eq:gammamu2})
\begin{align}
    (\hat{\Gamma}^+_j)^\dag&\propto \hat{\Gamma}_j^- \ .\label{eq:gamma_adjoint}
\end{align}
In contrast, for $\sigma=-1$, the only right-localized boundstate with positive eigenvalue is the quasi{\it particle}
\begin{align}
    \hat{c}_{R,-}^\dag.
\end{align}
Again, the reversion quantum number for this boundstate satisfies $\mu=+1$, which follows from direct inspection of \eqref{eq:text1}. On the other hand, on the left-hand-side of the graph, the positive-energy excitation
\begin{align}
    \hat{c}_{L,\pm}^\dag
\end{align}
has reversion number $\mu = -1$ (c.f. \eqref{eq:text2}). Crucially, this quantum number is insensitive to the value of $\sigma$, as $\sigma$ does not show up in the left-boundary condition (c.f. \eqref{eq:bcs_L}). In summary, therefore, when $\sigma=-1$, the left- and right-boundstates with positive energy have differing reversion quantum number $\mu$, and thus cannot hybridize. In contrast, when $\sigma=+1$, the left- and right-boundstates with positive energy are both {\it antisymmetric} ($\mu=-1$) under reversion, and can thus hybridize. The analysis for the case of the negative energy crossing proceeds in the exact same fashion. \\

\subsubsection{Location of crossings, and scaling of the gap}

We now derive the conditions for the existence of a spin glass bottleneck in the Frustrated Ring annealing schedule, as well as compute its location $B\equiv B_b$. Furthermore, we analytically calculate the scaling of the gap at the bottleneck location. To obtain the crossing point of the left- and right-bound state energies,
we set them equal to each other, i.e. $\epsilon_{\sigma\mu}^R= \epsilon^L_{\mu}\equiv
\epsilon_b$. We will deal with both crossings at the same time: from the discussion in the previous paragraph, we have that the left-localized boundstate involved in the positive-energy crossing has $\mu = -1$ for both the even and odd-crossing, so the left-localized boundstate involved in the positive-energy crossing is always
\begin{align}
    \psi^L_-.
\end{align}
For the right-localized boundstate involved in the positive-energy crossing, from the discussion in the previous paragraph, the value of $\mu$ depends on the fermion parity $\sigma$. However, we can compute the product $\sigma\mu =-1$, which is the same in both cases. Therefore, the right-localized boundstate involved in the positive-energy crossing is
\begin{align}
    \psi^R_-.
\end{align}
Furthermore, we note that these states have the same energy $\epsilon\equiv \epsilon_b$, so by the dispersion relation \eqref{eq:dispersion}, we have that these crossing states have equal and opposite $\kappa\equiv \pm \kappa_b$. Therefore, by \eqref{eq:lambda}, the states also have equal and opposite $\lambda\equiv \pm \lambda_b$. Thus, at the crossing, these states therefore satisfy the boundary conditions
\begin{align}
   B_be^{\lambda_b} &= \epsilon_b -J_R \ ,\\
   B_be^{\lambda_b} &= \epsilon_b -\frac{J_L^2}{\epsilon_b +B_b} \ ,
\end{align}
which yield the identity
\begin{align}
    \epsilon_b+B_b=\frac{J_L^2}{J_R}.\label{eq:penult1}
\end{align}
Combining this with one of the bulk equation \eqref{eq:bulk_eq1} yields
\begin{align}
    e^{\kappa_b-\lambda_b}&=\frac{J_R}{J}.\label{eq:penult2}
\end{align}
Lastly, we can also consider using the bulk equation \eqref{eq:penult3}. The triplet of equations \eqref{eq:penult1}, \eqref{eq:penult2} and \eqref{eq:penult3}, considered together, yield the bottleneck location, as well as the crossing energy (e.g., using Solve in {\it Mathematica}):
\begin{align}
    B_b&=\frac{1}{J_R}\frac{(J^2-J_L^2)(J_L^2-J_R^2)}{J_R^2+J^2-2J^2_L},~~~\epsilon_b= \frac{1}{J_R}\frac{J_R^2J^2-J_L^4}{J_R^2+J^2-2J_L^2}.
\end{align}
Note that $JJ_R>J_L^2$  is thus a necessary condition for the crossing point to
exist. Also, crucially, the bottleneck location is independent of $\sigma$, and thus happens in the same location for both the even- and odd-fermion sectors. Using expressions for $B_b$ and $\epsilon_b$ to solve for $\cosh
\kappa_b$, we obtain
\begin{equation}
\kappa_b =\log\frac{J_R(J^2 -J_L^2)}{J(J_L^2 - J_R^2)} \ ,
\end{equation}
which determines the scaling of the hybridization of the boundstates, $\Delta_\text{min}
\sim \mO(e^{-\kappa_b\mN})$. The
location and the value of the gap can be tuned by adjusting the parameters
of the model.

\section{Field-theoretic calculation of the tunneling matrix elements} \label{app:calc}

\subsubsection{Relation to the corresponding ferromagnetic problem}
\begin{figure}
        \includegraphics[width=\columnwidth]{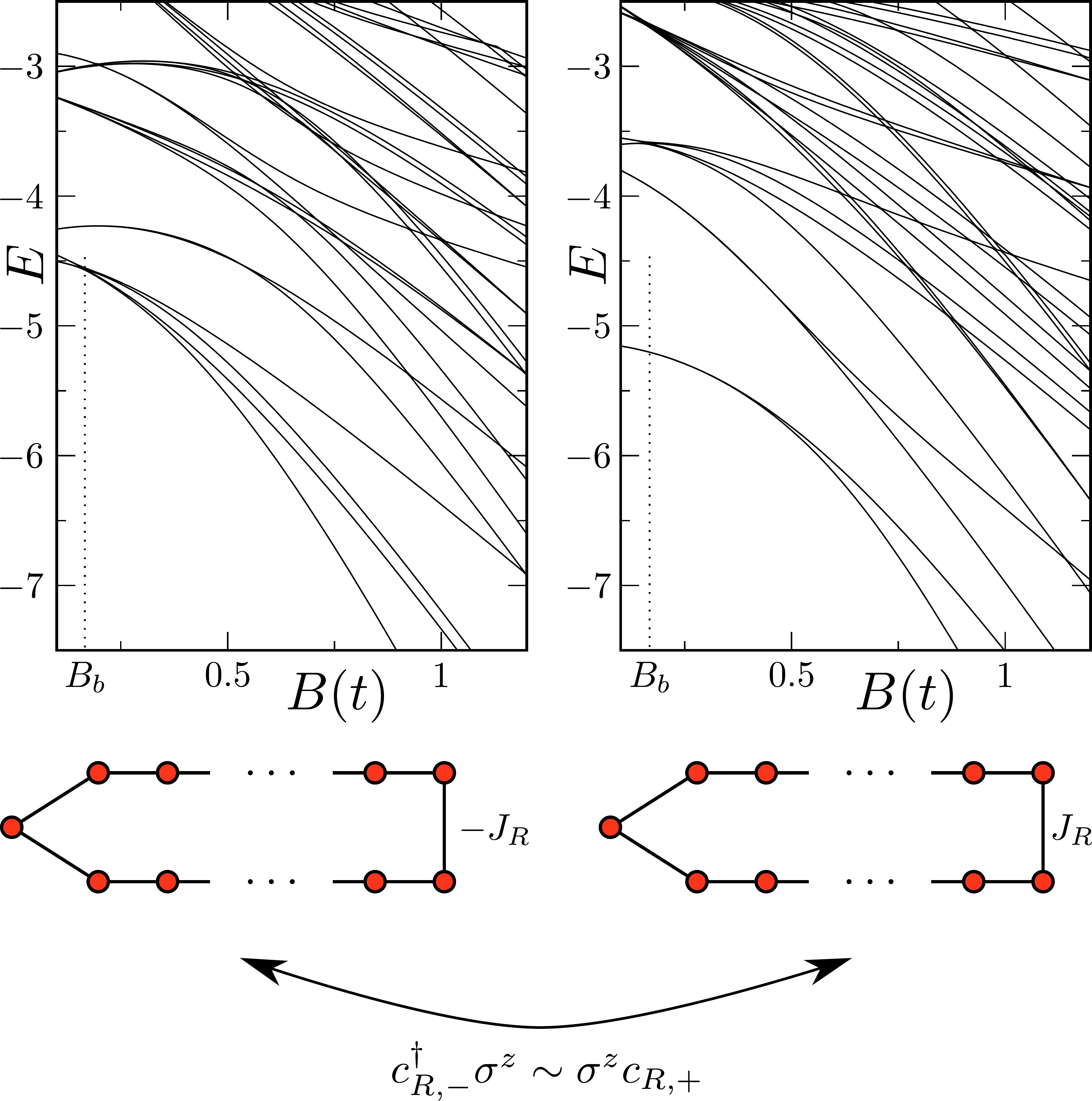}
        \caption{{\bf ``Unwinding'' the Frustrated Ring}. Passing an excitation from the left to the right of the qubit noise operator restores parity. More specifically, it removes frustration from the relevant matrix element, effectively restoring the gap. The left panel shows the spectrum of original Frustrated Ring. The bottleneck is removed by flipping the sign of $J_\mN$, as shown in the right panel. See text for details.}
        \label{fig:universality}
\end{figure}

We now turn to the relations used in the text, which transform
a calculation in a transverse-field Ising spin glass to one in a corresponding ferromagnetic
model $\widetilde{H}$. We then take advantage of this transformation to calculate the tunneling rates exactly via the quantum-classical correspondence. Indeed, recall that the transverse-field Ising chain maps to two free
fermion models $H^\pm$ (c.f. \eqref{eq:fmodel_pair_of_models}) (Ramond/Neveu-Schwarz), leading to an unphysical
doubling of the number of eigenstates. In our calculations up to this point, we have only considered the low-energy states
\begin{alignat}{4}
& \ket{\Psi_R^-} &&= c_{R,-}^\dag \ket{\Omega_-} \ , \quad &&\ket{\Psi_L^-}
&&= c_{L,-}^\dag \ket{\Omega_-} \ , \\ & \ket{\Psi_R^+} &&= \ket{\Omega_-} \ ,
\quad 
&&\ket{\Psi_L^+} &&= c_{L,+}^\dag c_{R,+}^\dag\ket{\Omega_+} \ ,
\end{alignat}
and have implicitly {\it discarded} the low-energy
states
\begin{alignat}{4}
&\ket{\widetilde{\Psi}_R^+} &&\equiv \ket{\Omega_-} \ , \quad &&
\ket{\widetilde{\Psi}_L^+} &&\equiv c_{R,-}^\dag c_{L,-}^\dag \ket{\Omega_-} \ , \\
&\ket{\widetilde{\Psi}_R^-} &&\equiv c_{R,+}^\dag \ket{\Omega_+}\ , \quad &&
\ket{\widetilde{\Psi}_L^-} &&\equiv c_{L,+}^\dag \ket{\Omega_+} \ ,
\end{alignat}
as they are not genuine eigenstates of the original
spin chain Hamiltonian (they have the wrong parity). However, utilizing the identity $H^\pm = \widetilde{H}^\mp$  ($\widetilde{H}$ is the ferromagnetic modification of the original spin glass Hamiltonian; see Eq. \eqref{eq:intertwining} in the main text, c.f. Figure \ref{fig:universality} for a depiction of this modification for the case of the Frustrated Ring), we can reinterpret the unphysical states of $H$ as {\it physical states} for $\widetilde{H}$:
\begin{alignat}{4}
&\ket{\widetilde{\Psi}_R^+} &&= \ket{\widetilde{\Omega}_+} \ , \quad &&
\ket{\widetilde{\Psi}_L^+} &&= \widetilde{c}_{R,+}^\dag \widetilde{c}_{L,-}^\dag \ket{\widetilde{\Omega}_+} \ , \label{eq:unphysical_1} \\
&\ket{\widetilde{\Psi}_R^-} &&= \widetilde{c}_{R,-}^\dag \ket{\widetilde{\Omega}_-}\ , \quad &&
\ket{\widetilde{\Psi}_L^-} &&= \widetilde{c}_{L,-}^\dag \ket{\widetilde{\Omega}_-} \ ,\label{eq:unphysical_2}
\end{alignat}
with similar relations for the remaining (i.e. bulk) eigenstates. Therefore, one can interpret the general situation in the following way: 
when one diagonalizes $H^+$ and $H^-$ in \eqref{eq:fmodel_pair_of_models}, yielding two
full sets of fermonic Fock states, half unphysical, it is actually valid to
say that one obtains two full sets of {\it physical} spin chain eigenstates,
one for the original spin chain, and the remaining set corresponding to the
eigenstates of its {\it frustration-free version}.\\ 

Now, consider the central problem encountered in the main text, namely, that of computing
\begin{align}
    \bra{\Psi_R^+}\sigma_j^z\ket{\Psi_R^-},~~~\bra{\Psi_L^+}\sigma_j^z\ket{\Psi_L^-}.
\end{align}
Naively, we can try to take advantage of the above observation and convert the matrix elements in our glassy problem $H$ into matrix elements in our ferromagnetic problem $\widetilde{H}$. Indeed, substituting, we get
\begin{align}
    \bra{\Psi_R^+}\sigma^z_j\ket{\Psi_R^-} &= \bra{\Omega_+}\sigma^z_jc_{R,-}^\dag\ket{\Omega_-} \nonumber\\
&=\bra{\widetilde{\Omega}_-}\sigma^z_j\widetilde{c}_{R,+}\ket{\widetilde{\Omega}_+},\\
    \bra{\Psi_L^+}\sigma^z_j\ket{\Psi_L^-} &= \bra{\Omega_+}c_{L,+}c_{R,+}\sigma^z_jc_{R,-}^\dag\ket{\Omega_-} \nonumber\\
&=\bra{\widetilde{\Omega}_-}\widetilde{c}_{L,-}\widetilde{c}_{R,-}\sigma^z_j\widetilde{c}_{R,+}^\dag\ket{\widetilde{\Omega}_+}.
\end{align}
Unfortunately, all of these matrix elements are between unphysical states, suggesting that one must pass excitations across the $\sigma^z$ operator, to amend the situation.

In general, since $\sigma^z$, when written-out in terms of $\hat{\gamma}$-matrices, is an element of the {\it Pin} group $\text{Pin}(2\mN)$ (see \cite{palmer_thermodynamic_2007} for a standard reference), passing $\sigma^z$ past a fermionic excitation will rotate that excitation by an orthogonal matrix \cite{palmer_thermodynamic_2007}, producing a new fermionic excitation:
\begin{align}
    \not{\psi'}&\equiv \sigma^z_j \not{\psi} \sigma^z_j \ ,
\end{align}
where
\begin{align}
    \psi' = R_j\psi \ ,
\end{align}
with $R_j\in O(2N)$ a rotation matrix. Utilizing this fact, we have
\begin{align}
    \bra{\Psi_R^+}\sigma^z_j\ket{\Psi_R^-} &= \bra{\Omega_+}\sigma^z_jc_{R,-}^\dag\ket{\Omega_-}\nonumber \\
&=\bra{\widetilde{\Omega}_-}R_j[c_{R,-}^\dag]\sigma^z_j\ket{\widetilde{\Omega}_+},\label{eq:R_j_eqn1}\\
    \bra{\Psi_L^+}\sigma^z_j\ket{\Psi_L^-} &= \bra{\Omega_+}c_{L,+}c_{R,+}\sigma^z_jc_{L,-}^\dag\ket{\Omega_-} \nonumber\\
&=\bra{\widetilde{\Omega}_-}\widetilde{c}_{L,-}\sigma^z_jR_j[c_{R,+}]\widetilde{c}_{L,+}^\dag\ket{\widetilde{\Omega}_+} \ \label{eq:R_j_eqn2}
\end{align}
where in deriving \eqref{eq:R_j_eqn2}, we have utilized the fact that $R_j^2=1_{2\mN\times 2\mN}$. Computing the rotation $\mathcal R_j\in O(2\mN)$ implemented by $\sigma^z_j$
is straightforward. In particular, by writing everything out in terms of $\hat{\gamma}$ matrices, we find, that, for $i<2j-1<2\mN-i$,
\begin{equation}
\mathcal R_j(\hat{\Gamma}_i^\mu)=-\hat{\Gamma}^{-\mu}_i,\\
\end{equation}
i.e. $\mathcal R_j$ flips the reversion-symmetry quantum number. From this,
due to the localized nature of the $R$-bound states, we obtain our desired
result:\\
\begin{align} 
\mathcal R_j(c_{R,\pm}^\dag)&=\mathcal R_j\bigg(\sum_i(\psi^R_\pm)_i \hat{\Gamma}_i^+\bigg) \nonumber\\
&=-\sum_i(\psi^R_\pm)_i\hat{\Gamma}_i^-+ \mO(e^{-\kappa j})\nonumber\\
&=~~~~-c_{R,\mp} + \mO(e^{-\kappa j}) \ ,\label{eq:frustrationeq}
\end{align}
which is essential to derive the main results of the paper. Substituting the identity \eqref{eq:frustrationeq} into Eq.'s (\ref{eq:R_j_eqn1}-\ref{eq:R_j_eqn2}) and comparing with Eqs. (\ref{eq:unphysical_1}-\ref{eq:unphysical_2}), we get the relations Eqs. \eqref{eq:intertwining_the_states} used in the main text:\\
\begin{eqnarray}
\bra{\Psi_R^+} \hat{\sigma}^z_j \ket{\Psi_R^-} &=&- \bra{\widetilde{\Psi}_R^-} \hat{\sigma}^z_j \ket{\widetilde{\Psi}_R^+} + \mO(e^{-\kappa |j-j_R|}),\ \label{eq:intertwining_the_states_R}\\
\bra{\Psi_L^+} \hat{\sigma}^z_j \ket{\Psi_L^-} &=&- \bra{\widetilde{\Psi}_L^-} \hat{\sigma}^z_j \ket{\widetilde{\Psi}_L^+} + \mO(e^{-\kappa |j-j_R|}). \ \label{eq:intertwining_the_states_L}\\
\nonumber
\end{eqnarray}

\subsubsection{Field-theoretic treatment of the ferromagnetic problem}
We now give exact expressions for the matrix elements given in Eqs. (\ref{eq:intertwining_the_states_R}-\ref{eq:intertwining_the_states_L}), involving eigenstates of the ferromagnetic Hamiltonian $\widetilde{H}$. We begin by calculating the matrix element corresponding to the ferromagnetic groundstate:
\begin{align}
    \bra{\widetilde{\Psi}_R^+}\sigma^z_j\ket{\widetilde{\Psi}_R^-} \ .
\end{align}
Since the frustration-free model is globally gapped in the spin glass (i.e. ferromagnetic) phase $B<B_c$, there are no crossings involving the groundstate manifold (See Figure \ref{fig:universality}), and 
\begin{align}
\ket{\Psi_{GS}^+} &\equiv \ket{\widetilde{\Psi}_R^+} \ , \label{eq:gs1}\\
\ket{\Psi_{GS}^-} &\equiv \ket{\widetilde{\Psi}_R^-} \ ,\label{eq:gs2}
\end{align}
span the degenerate ground state manifold in the thermodynamic limit of infinite-chain length $\mN\to \infty$. We begin by showing that the quantity we wish to compute, is really the spontaneous magnetization of our quantum spin chain, in disguise. Since the model is completely ferromagnetic (as we have removed frustration), if we perturb the model with an appropriately-aligned longitudinal field, i.e. define 
\begin{align}
\widetilde{H}_0(h)\equiv \widetilde{H}_0+\frac{h}{\mN}\sum_{j=1}^{\mN} \hat{\sigma}^z_j,
\end{align}
then the groundstate degeneracy is broken. Here, the external field is scaled as $\sim O(1/\mN)$ in order to ensure the perturbation is bounded in the thermodynamic limit $\mN \to\infty$. Since the perturbation mixes fermion parity, simple degenerate perturbation theory in the groundstate manifold Eqs. (\ref{eq:gs1}-\ref{eq:gs2}) yields that the perturbed ground-state, in the limit $h\to 0^+$, is 
\begin{align}
|\Psi_{GS}^{(0)}\rangle&\equiv \lim_{h\to 0^+}\lim_{\mN\to\infty}|\Psi_{GS}(h)\rangle\\
&= \lim_{\mN\to\infty}\frac{1}{\sqrt{2}}(\ket{\widetilde{\Psi}_R^+} +\ket{\widetilde{\Psi}_R^-} ).\label{eq:B5}
\end{align}
The reason we consider this perturbation of the frustration-free model is because this allows us to relate our transition matrix element to the {\it spontaneous magnetization} of this model. Indeed, since $\hat{\sigma}^z_j$ mixes fermion parity,
\begin{align}
&\lim_{\mN \to \infty}\bra{\widetilde{\Psi}_R^+} \hat{\sigma}^z_j\ket{\widetilde{\Psi}_R^-}\nonumber\\
&=\lim_{\mN\to \infty} \frac{1}{2}(\bra{\widetilde{\Psi}_R^+} +\bra{\widetilde{\Psi}_R^-} )\hat{\sigma}^z_j(\ket{\widetilde{\Psi}_R^+} +\ket{\widetilde{\Psi}_R^-} )\nonumber\\
&~~~~~~~~~~~~~~~~~~~~~~~~~~~~~~~~~~~~=\bra{\Psi_{GS}^{(0)}}\hat{\sigma}^z_j \ket{\Psi_{GS}^{(0)}}.\\
\nonumber
\end{align}
To compute this spontaneous magnetization, we write
\begin{align}
\lim_{\mN\to\infty}\bra{\widetilde{\Psi}_R^+} \hat{\sigma}^z_j\ket{\widetilde{\Psi}_R^-}&=\lim_{h\to 0^+}\lim_{\mN\to \infty}\bra{\Psi_{GS}(h)}\hat{\sigma}^z_j \ket{\Psi_{GS}(h)}\nonumber\\
&=\lim_{h\to 0^+}\lim_{\mN\to \infty}\lim_{T\to 0}\text{Tr}[e^{- \widetilde{H}_0(h)/T}\hat{\sigma}^z_j ].
\end{align}
This allows us to probe this matrix element using the quantum-classical correspondence: we begin by defining a partition function via
\begin{align}
Z_{\mN,h}&\equiv\lim_{T\to 0} \text{Tr}[e^{- \widetilde{H}_0(h)/T}].
\end{align}
As is standard, to obtain the correspondence with a classical model, we apply the Suzuki-Trotter transformation with a time-step $\tau_c>0$ to the partition function, producing a family of effective actions $\{S[\tau_c]\}_{\tau_c>0}$ describing (classical) stochastic fluctuations of an Ising spin system on a cylindrical spacetime lattice. After a straightforward manipulation, one gets \cite{sachdev_quantum_2000}:
\begin{equation}
Z_{\mN,h}= \sum_{\{s_{j\tau}=\pm
1\}}e^{-S[\tau_c,h]} \ .
\end{equation}
Here, $\tau_c \ll 1$ is a UV
cutoff defining a non-perturbative renormalization group flow in imaginary
time \cite{costello_renormalization_2011}. For small values of the UV cutoff, the action of the statistical field theory simplifies to \cite{sachdev_quantum_2000}:
\begin{align}
S[\tau_c,h] &\underset{\tau_c\to 0}{\sim}\sum_{j, \, \tau\in \tau_c\mathbb{Z}}(J_j[\tau_c]
s_{j\tau} s_{j+1,\tau}+J_\perp[\tau_c] s_{j\tau}s_{j,\tau +\tau_c}\nonumber\\
&~~~~~~~~~~~~~~~~~~~~~~~~+H[\tau_c] s_{j,\tau}) \ ,\label{eq:action}
\end{align}
where the coupling constants in our theory have the following dependence on the cutoff (for small values of the cutoff):
\begin{align}
    J_\perp[\tau_c]&\equiv\ln \tanh (B\tau_c),\nonumber~~~~~~~~    H[\tau_c]\equiv \tau_c h/\mN,\\
    J_j[\tau_c]&\equiv \tau_c J_j.\nonumber
\end{align}
Thus, we have
\begin{align}
\lim_{\mN\to\infty}\bra{\widetilde{\Psi}_R^+} \hat{\sigma}^z_j\ket{\widetilde{\Psi}_R^-}&=\lim_{\alpha\to 0^+} \lim_{\substack{\mN\to\infty, \\H=\alpha/\mN}}\lim_{\mN_\perp\to\infty}\langle s_{j,\tau}\rangle_{H} \ ,\label{eq:yangmag}
\end{align}
which is exactly {\it Yang's definition} (as reviewed in \cite{schultz_two-dimensional_1964}) of the spontaneous magnetization of the two-dimensional {\it classical Ising model} (\ref{eq:action}). Following \cite{schultz_two-dimensional_1964}, we denote this with the shorthand $M_j$, so that Eq. (\ref{eq:yangmag}) is equivalently stated as
\begin{align}
\lim_{\mN\to\infty}\bra{\widetilde{\Psi}_R^+} \hat{\sigma}^z_j\ket{\widetilde{\Psi}_R^-}&=M_j \ ,
\end{align}
where it is understood that we are applying Yang's definition of the spontaneous magnetization to the action Eq. (\ref{eq:action}), which lacks translational symmetry in the spatial direction.

In the case that the weights in our MAXCUT problem are uniform in absolute value, i.e. when $|J_j|\equiv J$ for all $j$, then the two-dimensional Ising model corresponding to the ferromagnet $\widetilde{H}$ is uniform, and we can cite Yang's result \cite{yang_spontaneous_1952} here for the exact matrix element, which is independent of $j$:
\begin{align}
    M_j &=M,~~~~M\equiv (1-k^{-2})^{1/8} \ .
\end{align}
Here, $k$ is called the {\it spectral parameter}, and has the following exact form \cite{palmer_thermodynamic_2007}:
\begin{align}
k\equiv \sinh J[\tau_c] \sinh  J_\perp[\tau_c] \ .
\end{align} 
In this (uniform) case, we can calculate the spectral parameter in the UV limit of our field theory, in which case we get the ratio $J/B$ coming from the quantum spin chain:
\begin{align}
k&=\sinh J[\tau_c]\sinh J_\perp[\tau_c]\nonumber\\
&\underset{\tau_c\to 0}{\sim}J[\tau_c] e^{-J_\perp[\tau_c]}\underset{\tau_c\to 0}{\sim}\frac{J}{B}.
\end{align}
In other words, in the uniform case $|J_j|\equiv \text{const}$, the matrix element in the spin chain comes out to, in the large-$\mN$ limit:
\begin{align}
\lim_{\mN\to\infty}\bra{\widetilde{\Psi}_R^+} \hat{\sigma}^z_j\ket{\widetilde{\Psi}_R^-}&= (1-k^{-2})^{1/8} \ ,
\end{align}
for all $B(t)<J$ (i.e. the ordered phase for the ferromagnetic problem). \\

\subsubsection{Transfer matrix calculation}

In the quantum spin glass problem that we consider, the couplings are non-uniform. However, the broken translation invariance in the spin chain is due to the modification of only $3$ couplers, namely the couplers
\begin{align}
    J_{n}=J_{n+1}&\equiv J_L,\nonumber\\
    J_{2n+1}&\equiv J_R.
\end{align}
(Note that here, we are using the values of these couplers in the ferromagnetic version $\widetilde{H}$ of our model). Due to the finite correlation length in the classical model, the effect of local changes to the coupling constants in the theory is washed-out in the thermodynamic limit, when we sum the spontaneous magnetization over all sites $j$. That is, we can expect the behavior
\begin{align}
     \sum_{j=1}^\mN M_j^2 \underset{\mN\to \infty}{\sim} \mN M^2.\label{eq:asymptote}
\end{align}
where, here, $M\equiv (1-k^{-2})^{1/8}$ is the result for the uniform chain. To demonstrate the asymptotic result Eq. (\ref{eq:asymptote}), we must demonstrate a {\it boundary effect} in the classical model. To do this, we calculate the local spontaneous magnetization using the row transfer matrix.\\

Indeed, consider computing the spontaneous magnetization (representing the tunneling matrix element in the quantum spin glass) using Yang's algorithm (as reviewed in \cite{schultz_two-dimensional_1964}), but now applied to the {\it row transfer matrix}, as opposed to the column transfer matrix:
\begin{align}
&M_\text{$j$} =\lim_{\alpha\to 0^+} \lim_{\mN\to\infty}\lim_{\substack{\mN_\perp\to\infty, \\H=\alpha/\mN_\perp}}\langle s_{j,\tau}\rangle_{H}\nonumber\\
&=\lim_{\alpha\to 0^+} \lim_{\mN\to\infty}\lim_{\substack{\mN_\perp\to\infty, \\H=\alpha/\mN_\perp}}\frac{\text{Tr}[T[J_1;H]\cdots \hat{\sigma}^z_\tau \cdots T[J_\mN;H]]}{\text{Tr}[T[J_1;H]\cdots T_\mN[J_\mN;H] ]}.\label{eq:row}
\end{align}
Here, the row transfer matrix is (see, e.g. \cite{palmer_thermodynamic_2007})
\begin{align}
    T[J_l;H]&\equiv  \bigg(\prod_{\tau}e^{J_\perp[\tau_c] \hat{\sigma}^z_{\tau}\hat{\sigma}^z_{\tau+1}+H\hat{\sigma}^z_\tau}\bigg)\nonumber\\
    &~~~\cdot\bigg(\prod_{\tau}e^{J_l[\tau_c]}(1+e^{-2J_l[\tau_c]}\hat{\sigma}^x_\tau)\bigg).
\end{align}
The spectral parameter for this transfer matrix when $H \equiv 0$ can be calculated, and comes out to
\begin{align}
    k_l&=\sinh J_l[\tau_c]\sinh J_\perp[\tau_c]\nonumber\\
&\underset{\tau_c\to 0}{\sim}J_l[\tau_c] e^{-J_\perp[\tau_c]}\underset{\tau_c\to 0}{\sim}\frac{J_l}{B}.
\end{align}
Note that, since the model is no longer translation-invariant, this spectral parameter is now dependent on the qubit location $l\in \{1,\cdots, \mN\}$. We can then proceed with the calculation, letting $d$ denote the distance between the site $j$ (where we are calculating the local spontaneous magnetization), and the nearest defect (e.g. the $J_L$ or $J_R$ coupler):
\begin{align}
    &M_\text{$j$}=\lim_{\alpha\to 0^+} \lim_{\mN\to\infty}\lim_{\substack{\mN_\perp\to\infty, \\H=\alpha/\mN_\perp}}\nonumber\\
    &\frac{\text{Tr}[T[J;H]^d\hat{\sigma}^z_\tau T[J;H]^{N-d}T[J_L;H]^2T[J;H]^NT[J_R;H]]}{\text{Tr}[T[J;H]^N T[J_L;H]^2T[J;H]^NT[J_R;H]]}\nonumber
\end{align}
with $N \equiv n-1$ equal to the bulk chain length. We can now see the emergence of a boundary effect in the classical model (and thus, by the quantum-classical correspondence, in the quantum ferromagnet $\widetilde{H}$ as well): letting $|\Psi_+\rangle$ denote the maximal eigenvector of $T[J;H]$, we have, by analogous arguments to Eqs. (\ref{eq:gs1}-\ref{eq:B5}), the following limiting behavior:\\
\begin{align}
    \lim_{\alpha\to 0^+}\lim_{\substack{\mN_\perp \to \infty\\ H=\alpha/\mN_\perp}}|\Psi_+\rangle&=\frac{1}{\sqrt{2}}(|\Psi^+_0\rangle +|\Psi^-_{k=0}\rangle) \ ,\label{eq:B22}
\end{align}
where here, $|\Psi_0^+\rangle$ and $|\Psi_{k=0}^-\rangle$ are the even- and odd-parity maximal-eigenvectors of $T[J;H\equiv 0]$ (following the notation of \cite{schultz_two-dimensional_1964}). Therefore, we can write
\begin{align}
    \lim_{\alpha\to 0^+}\lim_{\substack{\mN_\perp\to \infty\\ H=\alpha/\mN_\perp}}T[J;H]^d&=|\Psi_+\rangle \langle \Psi_+|+O(e^{-d\Delta}) \ , \label{eq:B23}
\end{align}
where, here, $\Delta$ is the spectral gap for the unperturbed transfer matrix $T[J;H\equiv 0]$, which sets the correlation length in the spatial direction of the lattice. Using Eqs. (\ref{eq:B22}-\ref{eq:B23}), we have
\begin{align}
    &M_\text{$j$}=\frac{\langle \Psi_+|\hat{\sigma}^z_\tau|\Psi_+\rangle}{\langle \Psi_+|\Psi_+\rangle}+ O(e^{-\Delta d}) \ ,
\end{align}
where we have used the fact that, for large $\mN$, $N-d\to \infty$, where $N\equiv n-1$ is the bulk chain length. According to \cite{schultz_two-dimensional_1964}, the leading-order term in the above expression comes out to\\
\begin{align}
    \frac{\langle \Psi_+|\hat{\sigma}^z_\tau|\Psi_+\rangle}{\langle \Psi_+|\Psi_+\rangle}&=(1-k_0^{-2})^{1/8},
\end{align}
where $k_0$ is the spectral parameter for the bulk transfer matrix $T[J;H\equiv 0]$, which is simply $J/B$. Therefore, in total, at a distance $d$ away from either the right- or left-end
of the chain,
\begin{equation}
\bra{\widetilde{\Psi}_R^-}\hat{\sigma}^z_j\ket{\widetilde{\Psi}_R^+} \underset{\mN \to \infty}{\sim}M + \mO(e^{-\Delta d}) \ .
\end{equation}
This bulk convergence behavior is confirmed by exact diagonalization with up to $\mN=23$ sites, see Figure \ref{fig:gapped} in the main text. The other tunneling matrix
\begin{align}
    \bra{\widetilde{\Psi}_L^+}\hat{\sigma}^z_j\ket{\widetilde{\Psi}_L^-} \label{eq:other}
\end{align}
can be computed similarly: one begins by noting that $|\Psi_L^+\rangle$ is related to $|\Psi_R^+\rangle$ by applying two quasiparticle
operators (and the same is true for the relationship between $|\Psi_L^-\rangle$ and $|\Psi_R^-\rangle$). Since these operators are localized at opposite locations of the graph, their product gets mapped,
under the Jordan-Wigner transformation, to a string of spin flips $\prod_j \sigma^x_j$ between the centers
$j_L$ and $j_R$ of the corresponding bound state wave functions. Therefore,
because conjugation by a product of spin flips only has the potential to flip the sign of
the magnetization, the tunneling form factors corresponding to \eqref{eq:other} has the exact same asymptotics
(a tunneling form factor is given by the squared absolute value of a matrix element of the
type considered above).

\end{document}